\documentclass[preprint,aps,showpacs,floatfix]{revtex4}
 \usepackage{epsfig}
\usepackage{axodraw}
\usepackage{graphics}
\usepackage{graphicx}
\usepackage{epsf}
\usepackage{amssymb}
\usepackage{amsmath}

\def\be{\begin{eqnarray} &&}

\def\ee{\end{eqnarray}}

\def\bew{\begin{widetext}}
\def\ew{\end{widetext}}

\def\Dslash{\raise.15ex\hbox{/}\kern-.7em D}
\def\Pslash{\raise.15ex\hbox{/}\kern-.7em P}

\newcommand{\bge}{\begin{equation}}
\newcommand{\ene}{\end{equation}}
\newcommand{\bea}{\begin{eqnarray}}
\newcommand{\eea}{\end{eqnarray}}
\newcommand{\bg}{\begin{eqnarray}}
\newcommand{\en}{\end{eqnarray}}

\def\qbar{\bar{q}}

\begin{document}

\title{
\vspace{-1cm} 
\begin{flushright}
    {\large LFTC-17-5/5 \vspace{1cm}}
\end{flushright}
Study of the in-medium nucleon electromagnetic form factors  
using a light-front nucleon wave function combined  
with the quark-meson coupling model}

\author{W.~R.~B.~de~Ara\'ujo$^a$, J.~P.~B.~C.~de~Melo$^b$, K.~Tsushima$^b $}
\affiliation{$^a$ Secretaria de Educa\c c\~ao do Estado de S\~ao Paulo,DE Norte 2,~S\~ao Paulo,~SP,~Brazil.}
\affiliation{$^b$  Laborat\'orio de F\'\i sica Te\'orica e Computacional, Universidade Cruzeiro do Sul, 
01506-000, S\~ao Paulo, SP, Brazil}
\date{\today}

\begin{abstract}
We study the nucleon electromagnetic (EM) form factors in symmetric nuclear matter 
as well as in vacuum within a light-front approach using the in-medium inputs  
calculated by the quark-meson coupling model.
The same in-medium quark properties are used as those  
used for the study of in-medium pion properties.
The zero of the proton EM form factor ratio in vacuum, 
the electric to magnetic form factor ratio $\mu_p G_{Ep}(Q^2)/G_{Mp}(Q^2)$  
($Q^2 = -q^2 > 0$ with $q$ being the four-momentum transfer),
is determined including the latest experimental data   
by implementing a hard constituent quark component in the nucleon wave function. 
A reasonable fit is achieved for the ratio $\mu_pG_{Ep}(Q^2)/G_{Mp}(Q^2)$ in 
vacuum, and we predict that the $Q_0^2$ value to cross the zero  
of the ratio to be about 15 GeV$^2$. 
In addition the double ratio data of the proton EM form factors 
in $^4$He and H nuclei, 
$[G^{^4{\rm He}}_{Ep}(Q^2)/G^{^4{\rm He}}_{Mp}(Q^2)]/[G^{^1{\rm H}}_{Ep}(Q^2)/G^{^1{\rm H}}_{Mp}(Q^2)]$,  
extracted by the polarized ($\vec{e}, e' \vec{p}$) 
scattering experiment on $^4$He at JLab, are well described. 
We also predict that the $Q_0^2$ value satisfying $\mu_pG_{Ep}(Q_0^2)/G_{Mp}(Q_0^2) = 0$  
in symmetric nuclear matter, shifts to a smaller value as increasing nuclear matter density,   
which reflects the facts that the faster falloff of $G_{Ep}(Q^2)$ as increasing $Q^2$   
and the increase of the proton mean-square charge radius.  
Furthermore, we calculate the neutron EM form factor double ratio 
in symmetric nuclear matter for $0.1 < Q^2 < 1.0$ GeV$^2$.  
The result shows that the neutron double ratio is enhanced 
relative to that in vacuum, while for the proton it is quenched, and  
agrees with an existing theoretical prediction.   
\end{abstract}
\pacs{24.85.+p, 21.65.-f, 14.20.Dh, 13.40.Gp, 12.39.-x}
\maketitle

\section{Introduction}
\label{intro}

One of the most challenging and exciting topics in hadronic and nuclear physics is 
how the hadron properties are modified in a nuclear medium (nuclear environment), 
and how such modifications can be measured in experiment.
Since hadrons are composed of quarks, antiquarks and gluons, it is natural to expect that 
the hadron internal structure is modified when they are immersed in a nuclear medium   
and in atomic nuclei~\cite{Brown,Hatsuda,Guichon,QMCreview,Hayano,Brooks,Krein:2017usp}.
At sufficiently high nuclear density and/or temperature, there is no doubt that 
the quark and gluon degrees of freedom are the correct degrees of freedom to describe 
the properties of hadrons according to quantum chromodynamics (QCD). 
On the other hand, it is also true that effective description of hadronic and nuclear processes 
is very successful by means of the meson and baryon degrees of freedom,  
especially in a lower energy and temperature region.
Although there is hope that lattice QCD simulation eventually can describe consistently 
the properties of hadrons in a nuclear medium as well as nucleus itself, 
the current status using physical pion mass value  
seems still difficult to get a reliable result   
at finite nuclear density~\cite{lattice1,lattice2,lattice3,lattice4}.

To understand the deep inelastic scattering (DIS) data at momentum transfer of several GeV, 
one certainly needs explicit quark degrees of freedom~\cite{Cardarelli1995,Araujo1995,Denig2013}.
In particular, the nuclear  European Muon Collaboration (EMC) 
effect~\cite{NEMC,Hen} has suggested the necessity of  
including the degrees of freedom beyond the traditional nucleon and mesons.
Furthermore, there is strong implication for the modifications of the bound proton 
electromagnetic (EM) form factors in the measurement of the double ratio  
of proton-recoil polarization transfer coefficients in ($\vec{e},e'\vec{p}$) scattering experiments 
on $^{16}$O and $^4$He nuclei at MAMI and JLab~\cite{Strauch1,Strauch2,Strauch3}.
It is also clear that the properties of bound neutron is modified in a nucleus,  
since it becomes stable, while the free neutron mean life 
is about 880 seconds due to the beta decay emitting a proton 
and an antineutrino~\cite{Hen,neutron1}.

However, it is very difficult to unambiguously separate and identify
the observed effects by the relevant degrees of freedom. 
In particular, to distinguish the possible in-medium modifications     
due to the nucleon internal structure change 
in a nuclear medium~\cite{Hatsuda,Guichon,QMCreview,Hayano,Brooks,Krein:2017usp,NEMC,Hen,Lu1,Lu2,
Frederico,Batista2002,Wilson1},  
from those due to the conventional many-body effects, such as the final state interactions 
and meson exchange effects described at the hadronic degrees of freedom~\cite{Schiavilla}.  
Such separation may only  be possible in a model dependent manner, 
since general experimental data involve all the effects simultaneously.
Thus, the interpretation of the modifications observed in experiments 
has not yet been established. 

In this article, we study the modifications of the nucleon EM form factors 
in symmetric nuclear matter, focusing on the internal structure change of nucleon.
Namely, we study them based on the property change of the light-flavor 
($u$ and $d$) quarks inside the nucleon, using the in-medium inputs 
calculated by the quark-meson coupling (QMC) model~\cite{Guichon,QMCreview}.
The purposes of the present study may be summarized as follows:  
(i) Since the interpretation of the medium modifications observed in 
experiments has not yet been established, one needs to study the issue by various 
different approaches/models to understand better, 
(ii) it is very interesting to study the effect of three-valence-quark-(spin)coupling  
to form the nucleon wave function, not by additive, independent quark models 
such models as in Refs.~\cite{Guichon,QMCreview,Lu1,Lu2,Batista2002}, 
(iii) a preliminary, similar study using the same model 
exists~\cite{Wilson1}, but now we have the updated data for the proton 
EM form factors, and can study with the improved parameters for the nucleon 
wave function, especially improving the high $Q^2$ region behavior, 
(iv) the in-medium inputs in Ref.~\cite{Wilson1} were adapted from 
Ref.~\cite{Frederico,Batista2002} that based on the relativistic harmonic oscillator confining potential, 
however, it turned out that such approach cannot describe well the properties of 
finite nuclei without introducing nonlinear meson interaction terms 
at the meson and nucleon level Lagrangian~\cite{Xing}. Thus we tempt to use 
the in-medium inputs from the QMC model, which have successfully been applied 
for studying various nuclear and hadronic reactions as well as the properties 
of finite (hyper)nuclei, and (v) we calculate the in-medium neutron 
EM form factor double ratio in addition to that of the proton, which 
was predicted in Ref.~\cite{Cloet} to be enhanced in medium contrary 
to the proton case, and demonstrate that our model result indeed agrees with the prediction.
A similar approach as in the present study was already applied for 
the study of pion properties in symmetric nuclear 
matter~\cite{pimedium1,pimedium2,pimedium3,pimedium4}.  
Although there may be possible to have alternative explanations based on traditional 
nuclear physics approaches, our interest of this study is on the 
internal structure change of nucleon in a nuclear medium. 

For this purpose, we rely on a light-front model of nucleon in vacuum, the 
``relativistic quark-spin coupling'' model, which was 
used for studying the nucleon EM form factors~\cite{afsbw,Wilson2} 
as well as the nucleon EM and axial-vector~\cite{Suisso} form factors 
in vacuum with some extensions including one of the present authors. 
Although we focus on the in-medium modifications of nucleon EM form factors 
in this study, the model could also describe reasonably well 
the axial-vector form factor and the coupling constant $g_A$ (obtained values 
$g_A = 1.09 - 1.29$) in vacuum with the quark mass values $330 - 380$ MeV~\cite{Suisso}. 
Note that, the present model corresponds to the parameter $\alpha = 1$ 
in the model of Ref.~\cite{Suisso}, and has an extra high-momentum component  
in the nucleon wave function. However, as studied in Refs.~\cite{afsbw,Wilson2}, 
the introduction of the high-momentum component does not destroy the achieved 
good feature of the model in the lower $Q^2$ region, such as $g_A$ at $Q^2 = 0$. 
For a smaller quark mass value such as 220 MeV to be used in this study, 
an exact calculation is planned be performed in the near future.
This model can keep close connection with covariant field theory,  
and perform a three-dimensional reduction for the photo-absorption amplitude 
with the nucleon projected on the null-plane, $x^+ = x^0 + x^3 = 0$.
After the three-dimensional reduction, one can introduce the nucleon light-front 
wave function in the two-loop momentum integrations. 
For studying the nucleon EM form factors, 
the ``triangle diagrams'' with an impulse approximation is used.
In Ref.~\cite{afsbw} the hard-scale component in the nucleon wave function 
was firstly introduced to improve the description of the zero of the 
proton EM form factor ratio, $\mu_p G_{Ep}(Q^2)/G_{Mp}(Q^2)$ 
($Q^2 = - q^2 > 0$, $q$ the four-momentum transfer). 
In addition a detailed study was made for the different quark-spin coupling 
effects in Refs.~\cite{afsbw,Wilson2}.
It turned out that the neutron electric form factor can strongly constrain 
the quark-spin coupling in the nucleon wave function, and the model preferred the scalar-pair 
coupling. Furthermore, to describe the zero of the proton EM form factor ratio, 
the introduction of the hard-scale component in the nucleon wave function was crucial. 
Thus, we use the two-scale model of the nucleon wave function in vacuum 
with the scalar-pair coupling, and study the medium modifications of the 
nucleon EM form factors, where the scalar-pair coupling means that, 
as will be given in Eq.~(\ref{lag}), the coupling between 
the (three quarks)-nucleon coupling is made by the Lorentz scalar. Other possibilities 
of the couplings were also studied in Refs.~\cite{afsbw,Wilson2,Suisso}. 
Although it is also very interesting to study 
the medium effects on the nucleon axial-vector form factor within the same model, 
this is planned to be made in the near future.

We predict the~$Q_0^2$ value to cross the zero  
of the ratio,  $\mu_p G_{Ep}(Q_0^2)/G_{Mp}(Q_0^2) = 0$,  
to be about 15 GeV$^2$. 
Furthermore, the double ratio data of the proton EM form factors 
in $^4$He and H nuclei, 
$[G^{^4{\rm He}}_{Ep}(Q^2)/G^{^4{\rm He}}_{Mp}(Q^2)]/[G^{^1{\rm H}}_{Ep}(Q^2)/G^{^1{\rm H}}_{Mp}(Q^2)]$, 
extracted by the polarized ($\vec{e}, e' \vec{p}$) 
scattering experiment on $^4$He at JLab, turn out to be well described. 
We also predict the $Q_0^2$ value of $\mu_pG_{Ep}(Q_0^2)/G_{Mp}(Q_0^2) = 0$  
in a nuclear medium, shifts to a smaller value as increasing nuclear matter density.

The organization of this article is as follows.
In Sec.~\ref{model}, we explain the relativistic quark-spin coupling 
model of nucleon, two scale-model as well as the nucleon EM form factors 
in a light-front approach, and present the nucleon EM form factors in vacuum. 
In Sec.~\ref{qmatter} we review the properties of nuclear medium necessary 
to study the in-medium modifications of the nucleon EM form factors, 
the QMC model, and discuss the in-medium inputs.
We present main results of this study, the in-medium nucleon EM 
form factors in Sec.~\ref{mediumffs}.
Finally, we give summary and discussions in Sec.~\ref{summary}.

\section{Nucleon electromagnetic form factors in vacuum}
\label{model}

Here, we briefly review a light-front approach for the nucleon structure, the relativistic 
quark-spin coupling model, and the two-scale model~\cite{afsbw}.
The effective Lagrangian for the quark-spin coupling in
the nucleon~\cite{afsbw,Wilson2,Suisso,Araujo2004}, accounts for calculating the static EM
observables with a totally symmetric momentum component of the nucleon 
wave function. However, the initial version of the model was necessary to be improved to
describe better the zero of 
$\mu_p G_{Ep}(Q^2)/G_{Mp}(Q^2)$~\cite{Jones2000,Brash2002,Gayou2002,Punjabi2005,Ron2007,Puckett2010,Ron2011,Puckett2012},  
namely, the position of the zero to be shifted to a larger $Q^2$. 
The effective Lagrangian for the three constituent quarks  
coupled to form the nucleon wave function is given by~\cite{afsbw,Wilson2,Suisso,Araujo2004},
\begin{eqnarray}
{\cal{L}}_{\mathrm{N-3q}}=m_N\epsilon^{lmn}
\overline{\Psi}_{(l)} i\tau _2\gamma _5\Psi_{(m)}^C\overline{\Psi}
_{(n)}\Psi _N + H.C. , \label{lag}
\end{eqnarray}
where $\tau _2$ is the Pauli matrix operating in isospin space, and the color indices are
$\{l,m,n\}$ with $\epsilon^{lmn}$ being the totally antisymmetric
tensor. The conjugate quark field is $\Psi^C=C
\overline{\Psi}^T $ with $C=i \gamma^2\gamma^0$, the charge
conjugation matrix, and $T$ stands for transposition. 
Quark flavor and spin quantum numbers are implicit in the quark filed operators, and 
they will be treated properly when one calculates relevant matrix elements 
as was done in Refs.~\cite{afsbw,Wilson2,Suisso,Araujo2004}, and partly 
shown in Appendix~\ref{appendix}.

The momentum scale of the nucleon wave function for Gaussian and
power law shapes with constituent up and down quark mass $m_q = 0.22$~GeV, was
found to be about 0.6 - 0.8 GeV from the fit to the nucleon magnetic moments 
and mean-square charge radii~\cite{afsbw}. 
(We note that the same value of the up and down quark constituent mass $m_q = 0.22$ GeV 
was also used for the studies of pion properties in vacuum~\cite{Pacheco1}, as well as  
in medium~\cite{pimedium1,pimedium2,pimedium3,pimedium4}.)
It turned out that the neutron electric form factor 
can constrain the relativistic quark-spin coupling scheme,  
and the scalar-pair coupling in the effective Lagrangian is preferred.

However, although the effective Lagrangian approach to the quark-spin
coupling allows a reasonable account for the static
nucleon EM observables with a totally symmetric
momentum component of the nucleon wave function, 
it has a too small momentum scale which leads to too small value for 
the zero of $\mu_p G_{Ep}(Q^2)/G_{Mp}(Q^2)$~\cite{afsbw,Wilson1}  
than that the experimental data 
imply~\cite{Jones2000,Brash2002,Gayou2002,Punjabi2005,Ron2007,Puckett2010,Ron2011,Puckett2012}.   

Therefore within this approach, one is led to introduce another
term in the momentum component of the nucleon wave function, 
which would represent a higher-momentum scale, to be able to describe better 
the zero of $\mu_p G_{Ep}(Q^2)/G_{Mp}(Q^2)$ without destroying the     
good description achieved in the lower momentum transfer 
region~\cite{afsbw,Wilson2}.

In some light-front models applied to 
mesons~\cite{pauli,tob01,sg,deMelo2004,deMelo2006,Pace2007,deMelo2009,Pace2010}, 
a high-momentum scale appears naturally associated with the short-range interaction 
between the constituent quarks. 
A reasonable description of the meson spectrum and pion
properties was achieved including a Dirac-delta interaction 
in the mass-squared operator~\cite{pauli,tob01,sg}, 
inspired by the hyperfine interaction from the effective one-gluon exchange 
between the constituent quarks~\cite{pauli,brodsky}. 
The model~\cite{sg} reveals some of the physics contained in the observation of 
the trajectories of mesons in the $(n,M^2)$-plane, 
that are almost linear~\cite{iachello,anisov}. 
The model naturally incorporates the small pion mass as a consequence of the short-range
attraction in the spin-zero channel, which is also responsible for
the pion and rho-meson mass splitting~\cite{sg}.

The short-range attractive part of the quark-quark interaction
which is presented in the Godfrey and Isgur model~\cite{god85}, 
generates a high-momentum component as well in the light-cone pion wave
function above the energy 1 GeV, and was successfully able to describe the
electroweak structure of pion~\cite{card}. Nonetheless, it was
pointed out that the existing electroweak data were not enough to
draw a definite conclusion about the presence of the 
hard constituent-quark components in the hadron wave function~\cite{ji}.
Recently, this discussion led to a new insight when the 
valence-quark light-cone momentum distribution was probed in the
experiment of diffractive dissociation of 500-GeV $\pi^-$ into
dijets~\cite{ashery}, which supports the importance of the
asymptotic part of the wave function~\cite{pqcd2}, 
and the presence of a high-momentum component in the pion wave 
function~\cite{tob01}.

Motivated by the above discussions which indicate the
necessity of a strong short-range attractive interaction in the
spin-zero channel and a high-momentum tail in the pion valence
component in the wave function, we introduce a high-momentum component in the
valence nucleon wave function. 
We study the role of this high-momentum component in the calculation of 
the nucleon EM form factors. 
Indeed, the quality of the model description including the recent data for 
$\mu_p G_{Ep}(Q^2)/G_{Mp}(Q^2)$, is improved substantially as we show later.

Thus, we use the ``two-scale model'', which includes the high-momentum component in the nucleon 
wave function in a light-front approach, in an effective 
Lagrangian with the spin coupling between the quarks Eq.~(\ref{lag}) in a scalar form. 
Furthermore, we choose a power-law form~\cite{bsch,brodsky} for the momentum
component of the nucleon wave function, 
\begin{eqnarray}
\Psi_{\mathrm{Power}} &=& N_{\mathrm{Power}}\left[(1+M^2_0/\beta^2)^{-p}
+\lambda (1+M^2_0/\beta_1^2)^{-p} \right]\ , 
\label{wf1}\\
\lambda &=& \left[ (1 + M_H^2/\beta_1^2)/(1 + M_H^2/\beta^2) \right]^p\ ,
\label{lambda}
\end{eqnarray}
which preserves the asymptotic behavior suggested by QCD.  
(See Eq.~(\ref{eqn:M0}) in Appendix A for the expression of $M^2_0$.)
The normalization constant $N_{\mathrm{Power}}$ above is determined by the proton charge. 
The characteristic momentum scales of the wave
function are represented by~$\beta$,~$\beta_1$~and~$M_H$, while~$M_0$ is the free mass 
of the three-quark system, and its explicit expression 
is also given in Ref.~\cite{afsbw}. 
The lower momentum scale is essentially 
determined by the nucleon static observables, while the higher one is
related with the zero of $G_{Ep}(Q^2)$. A possible definition of the
high-momentum scale brought by Eq.~(\ref{wf1}) is the value of
$M_0$ at which the two terms are equal, therefore one easily gets, 
\begin{eqnarray}
\beta_H = \beta\beta_1 \left( \frac{1-|\lambda|^{\frac{1}{3}}}
{\beta_1^2|\lambda|^{\frac{1}{3}}-\beta^2} \right)^{\frac{1}{2}}~. 
\label{bh}
\end{eqnarray}
We stress that this value should be interpreted as a guiding
reference. Note that the asymptotic behavior of
Eq.~(\ref{wf1}) does not depend on the parameters.  
The totally symmetric forms of Eq.~(\ref{wf1}), due to the relativistic 
spin-coupling coefficients which depend on momentum, effectively lead
to the breaking of the SU(6) flavor symmetry as discussed in
Ref.~\cite{simula00}.

The falloff based on perturbative QCD arguments for the power-low, 
has a value of $p=3.5$ in Eq.~(\ref{wf1})~\cite{bsch,brodsky}. 
From the point of view of the static electroweak observables, the value of
$p$ does not present an independent feature, once one static
observable is fitted. Namely, the other parameters in Eq.~(\ref{wf1})  
are strongly correlated, as long as $p>2$ ~\cite{afsbw,bsch}. 
In this study, we choose $p=3$.

The light-front formulation of the nucleon electroweak form
factors in Ref.~\cite{afsbw} uses the effective Lagrangian 
Eq.~(\ref{lag}), to construct the coupling of the quark spin in the
valence component of the nucleon wave function. The form factor
calculation is made by an impulse approximation defined within
a covariant field theory. The nucleon virtual photon absorption
amplitude is projected on the three-dimensional hypersurface,
$x^+=x^0+x^3=0$ (see, e.g., Ref.~\cite{karmanov}).

The elimination of the relative light-front time between the particles
in favor of the global time propagation~\cite{sales00,tob92,Ji:1998hx}, 
comes from the analytical integration in the individual light-front energies
($k^-=k^0-k^3$) in the two-loop amplitude.  Then, the momentum
component of the nucleon light-front wave function is introduced
into the remaining form of the two-loop three-dimensional momentum
integrations which define the matrix elements of the electroweak
currents~\cite{afsbw,tob92,pach99}.

The plus component of the nucleon EM current ($J^+_N=J_N^0+J_N^3$)
for momentum transfers satisfying the Drell-Yan condition
$q^+=q^0+q^3=0$, is used to calculate the EM form
factors. The contribution of the Z-diagram is minimized in a
Drell-Yan frame, while the wave function contribution to the current is 
maximized~\cite{karmanov,tob92,pach99,brodsky,ji00}. 
We use the Breit-frame, where the four-momentum transfer $q=(0,\vec
q_\perp,0)$ is such that $(q^+=0)$ and $\vec q_\perp=(q^1,q^2)$,
satisfying the Drell-Yan condition.

The nucleon EM form factors are calculated with the
matrix elements of the current $J^+_N(Q^2)$ in the light-front
spinor basis in the Breit-frame with the Drell-Yan condition~\cite{chung,afsbw}. 
The Dirac and Pauli form factors are respectively given by, 
\begin{eqnarray}
F_{1N}(Q^2)&=&\frac{1}{\sqrt{1+\eta}}\langle N \uparrow|J^+_N(Q^2)| N \uparrow\rangle
\ , \nonumber \\
F_{2N}(Q^2)&=&\frac{1}{\sqrt{\eta}\sqrt{1+\eta}}\langle N \uparrow|J^+_N(Q^2)| N \downarrow\rangle
\ , \label{jp}
\end{eqnarray}
where $\eta=Q^2/4m_N$. The momentum transfer in the Breit-frame
is chosen along the x-direction, i.e., $\vec{q_\perp} = (\sqrt{Q^2},0)$.

The nucleon electric and magnetic form factors (Sachs form
factors) are given by:
\begin{eqnarray}
G_{EN}(Q^2)&=& F_{1N}(Q^2)-\frac{Q^2}{4m_N^2}F_{2N}(Q^2) \ ,
\nonumber \\
G_{MN}(Q^2)&=& F_{1N}(Q^2)+F_{2N}(Q^2) \ ,
\end{eqnarray}
with $N = \ p$ or $n$. $\mu_N = G_{MN}(0)$ is the magnetic moment
and $\kappa_N=F_{2N}(0)$ is the anomalous one. 
The nucleon mean-square charge radius $r^2_N$ is calculated by 
$r^2_N \equiv <r^2_N> = -6\frac{dG_{EN}(Q^2)}{dQ^2}|_{Q^2=0}$.

\begin{figure}[t]
\begin{center}
\vspace{5cm} \centerline{\
\begin{picture}(330,130)(-5,-130)
\GOval(30,25)(15,5)(0){.5} \GOval(120,25)(15,5)(0){.5}
\Vertex(80,70){4.0} \put(110,12){\makebox(0,0)[br]{$(2)$}}
\put(110,42){\makebox(0,0)[br]{$(1)$}}
\put(110,72){\makebox(0,0)[br]{$(3)$}} \Line(30,10)(120,10)
\Line(30,40)(120,40) \Line(30,70)(120,70)
\put(85,-5){\makebox(0,0)[br]{$(1a)$}}
\GOval(150,25)(15,5)(0){.5} \GOval(240,55)(15,5)(0){.5}
\Vertex(200,70){4.0} \Line(150,10)(240,10) \Line(150,40)(240,40)
\Line(150,70)(240,70) \put(210,-5){\makebox(0,0)[br]{$(1b)$}}
\GOval(30,-75)(30,5)(0){.5} \GOval(120,-60)(15,5)(0){.5}
\Vertex(80,-45){4.0} \Line(30,-105)(120,-105)
\Line(15,-75)(120,-75) \Line(30,-45)(120,-45)
\put(85,-120){\makebox(0,0)[br]{$(1c)$}}
\GOval(150,-60)(15,5)(0){.5} \GOval(240,-60)(15,5)(0){.5}
\Vertex(200,-45){4.0} \Line(150,-105)(240,-105)
\Line(150,-75)(240,-75) \Line(150,-45)(240,-45)
\put(210,-120){\makebox(0,0)[br]{$(1d)$}}
\end{picture}
}
\end{center}
\caption{Diagrammatic representation of the nucleon
photo-absorption amplitude. The gray blob represents the spin
invariant for the coupled quark pair in the effective Lagrangian 
Eq.~(\ref{lag}). The small filled circle attached to the quark line represents
the action of the EM current operator.} 
\label{fig1}
\end{figure}
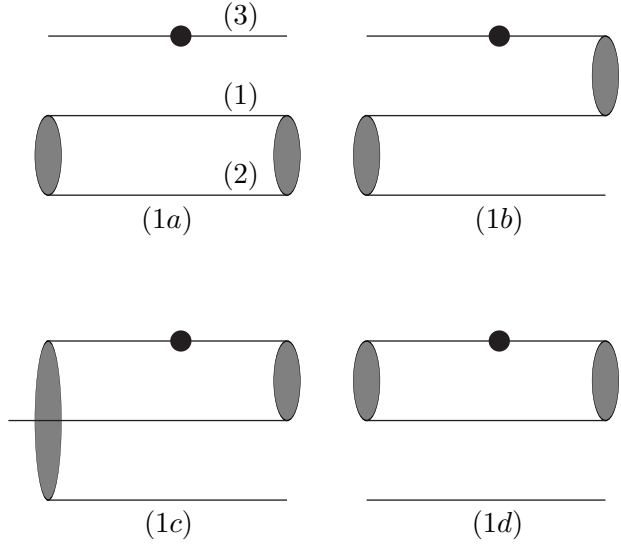

The microscopic matrix elements of the nucleon EM
current are derived from the effective Lagrangian Eq.~(\ref{lag}),
within the light-front impulse approximation which is represented
by four three-dimensional two-loop diagrams in Fig.~\ref{fig1}~\cite{afsbw}. 
The diagrams embody the antisymmetrization of the quark state in the wave function. 
In all diagrams in Fig.~\ref{fig1}, the quark 3 (the quark line with a small filled circle), 
is the quark which absorbs the momentum transfer by a photon. 
Figure (1a) [to be denoted by $J^+_{aN}$], defines the spin operator and represents the 
coupling between the quarks 1 and 2, while Fig. (1b) [to be denoted by $J^+_{bN}$], 
the coupled quarks in the nucleon initial state are the pair (1-3),  
and the coupled quarks in the nucleon final state are the  pair (1-2). 
This current $J^+_{bN}$ represented by diagram (1b) should be multiplied 
by a factor 4 --- a factor 2 comes from the exchange between the quanks 1 and 2, 
which are indistinguishable by this exchange, and the other factor 2 comes from 
the exchange between the pairs in the initial and final nucleon states, 
due the transformation by the time reversal and parity. 
The process represented by figure (1c) [to be denoted by $J^+_{cN}$], where, in the initial state  
the coupled quarks are pair (1-3) and in the final state the coupled quarks are 
pair (2-3). This current should be multiplied by a factor 2, since 
the quarks 1 and 2 can be exchanged. 
The process represented by diagram (1d) [to be denoted by $J^+_{dN}$], corresponds 
to the process in which a photon is absorbed by the diquark formed with the quarks 1 and 3, 
while the quark 2 is the spectator. In this case we must multiply a factor 2 
by possible exchange of the quarks 1 and 2. Therefore, 
the microscopic EM current operator for the nucleon depicted 
in Fig.~\ref{fig1} is given by, 
\begin{eqnarray}
J^+_N(Q^2)=J^+_{aN}(Q^2)+4J^+_{bN}(Q^2)+2J^+_{cN}(Q^2)+2J^+_{dN}(Q^2), 
\label{mjp}
\end{eqnarray}
with the factors explained. The explicit expressions 
and derivations made in Ref.~\cite{afsbw} are also summarized in Appendix A.

In this work the effective Lagrangian Eq.~(\ref{lag}), is a scalar coupling that
corresponds to the spin-coupling coefficients in which the Melosh
rotations of the quark spin have the arguments defined by the
kinematical momentum of the quarks in pair, and in the nucleon rest
frame constrained by the total momentum~\cite{araujo99,Araujo2006}, while in
the Bakamjian-Thomas construction  the argument of the Melosh
rotations are defined in the rest frame  of three free 
particles (constituent quarks).

The model of the nucleon adopted here assumes the dominance of the valence component, 
and the pion (meson) cloud effects, which are known to be important to simultaneously 
describe well the proton and neutron EM form 
factors e.g., in the cloudy bag model~\cite{PionCloud1}, 
light-front treatment~\cite{PionCloud2} and diquark approach~\cite{PionCloud3}, 
are not included explicitly. Within this approach, the results are strongly constrained, 
and the general features found in our calculations are rather  
independent of the detailed shape of the wave function, 
but depend on the momentum scales in Eq.~(\ref{wf1}). 
In the numerical evaluation of the form factors, we use a constituent
quark mass value of $m_q = 0.22$~GeV~\cite{afsbw,card} as mentioned. 
This value was also used in the study of pion properties in 
vacuum as well as in medium~\cite{pimedium1,pimedium2,pimedium3,pimedium4} 
with a light-front constituent quark model.
In addition, the model has three fitting parameters, 
the momentum scales $\beta$ and $\beta_1$, and the relative weight $\lambda$, 
or $M_H$ (see Eqs.~(\ref{wf1}) and~(\ref{lambda})).
We have selected two parameter sets for the present two-scale model, 
which will be denoted by ``set I'' and ``set II'',  
reproducing the proton magnetic moment ($\mu_p$), 
and neutron magnetic moment ($\mu_n$), respectively,  
as well as the zero of proton electric to magnetic form factor ratio $Q^2_0$  
for $\mu_p G_{Ep}(Q^2_0) / G_{Mp} (Q^2_0) = 0$, or $G_{Ep}(Q^2_0) = 0$.
In Tab.~\ref{tab1} we summarize the model parameters for the set I and set II, 
some nucleon static observables calculated, and the zero,~$Q^2_0$,~of~$G_{Ep}(Q^2_0) = 0$. 
\begin{table}[htbp]
\caption{Nucleon EM static observables and the zero of
$G_{Ep}(Q^2_0) = 0$, $Q^2_0$, for the two-scale models with the two sets of the parameters, 
set I and set II. The momentum-scale parameters, $\beta, \beta_1$ and $M_H$ 
in Eqs.~(\ref{wf1}) and~(\ref{lambda}), are given in the second, third and fourth
columns, respectively. The proton [neutron] magnetic moment $\mu_p$ [$\mu_n$] 
(in nuclear magneton) and proton root-mean-square charge radius $r_p \equiv <r^2_p>^{1/2}$ 
[neutron mean-square charge radius $r^2_n$], 
are given in fifth [seventh] and sixth [eighth] columns respectively. 
The zero, $Q^2_0$ value for $G_{Ep}(Q^2_0) = 0$, 
is given in the last column.
~See Tab.~\ref{table2}, for the experimental values.
\label{tab1}
}
\begin{center}
\begin{tabular}{|l|c|c|c|c|c|c|c|c|c|}
\hline 
\hline
 & $\beta$~(GeV) & $\beta_1$~(GeV) & $M_{H}$~(GeV) &$\mu_p$ &~$r_p$~(fm) & $\mu_n$&
$r^2_n$~(fm)$^2$ & $Q_0^2$~GeV$^2$ \\
\hline
Set I & 0.676   & 5.72 & 4.79  & 2.74   &~0.80  & -1.52 & -0.07 & 8.27  \\
Set II & 0.396    & 10.56 &5.92  & 3.05 &~0.94  & -1.88 & -0.06 & 15.12  \\
\hline
\hline 
\end{tabular}
\end{center}
\end{table}

\begin{table}[htbp]
\caption{Experimental values for some nucleon observables.}
\label{table2}
\begin{center}
\begin{tabular}{|l|l|l|l|l|}
\hline
\hline 
Ref.                    &  $\mu_p~[\mu_N]$ &~$\sqrt{r^2_p}~[fm]$&~$\mu_n~[\mu_N]$ &~$r^2_n~[fm^2]$  \\
\hline   
\cite{PDG2016,Mohr2015} &~$2.792847351\pm10^{-9}$  &~$0.8751\pm0.000061$   &   & \\
\cite{Bernauer2010}     &~                         &~$0.879\pm0.0008$      &   &  \\ 
\cite{PDG2016}          &                          &                       
&~$-1.9130427\pm5.10^{-7}$ &~$-0.1161\pm0.0022$  
\\
\hline
\hline 
\end{tabular}
\end{center}
\end{table}
We remind that, a single-scale nucleon light-front wave function, Gaussian or power-law,
with the proton or neutron magnetic moment fitted, is known to
give a reasonable proton charge radius, due to the strong
correlation between these observables ~\cite{bsch,brodsky,afsbw}.
However, the zero of $G_{Ep}(Q^2)$, $Q_0^2$, appears at too 
small values in the range 3-4~GeV$^2$. 
When we attempt to fit $Q_0^2$ to the values around or larger than 8~GeV$^2$ 
by increasing the momentum scale in the Gaussian and power-law nucleon wave functions 
of the one-scale model, we find too small proton size, and consequently bad magnetic
moment values. Thus, the facts leave us no room for improving the one-scale-based models.
However, by introducing a two-scale, namely, power-law high-momentum component in the
wave function with the scalar coupling, we are able 
to get a reasonable description of both the nucleon static
observables and the zero of $G_{Ep}(Q^2)$,~as shown in Tab.~\ref{tab1}.  
The scalar coupling provides the best agreement with the neutron
mean-square charge radius, when the neutron magnetic moment is fitted ~\cite{afsbw}. 
Note that, the high-momentum scale $M_H\sim 7.6$~GeV, should be understood 
as a reference value, and we point out that we cannot exclude completely the 
lower values for $M_H$, as one can see the  parameterization, $M_H = 4.79$ GeV.
Using these two parameter sets, set I and set II, we study the nucleon 
EM form factors in vacuum and in symmetric nuclear matter.

\begin{figure}[htbp]
\includegraphics[scale=0.44,angle=0]{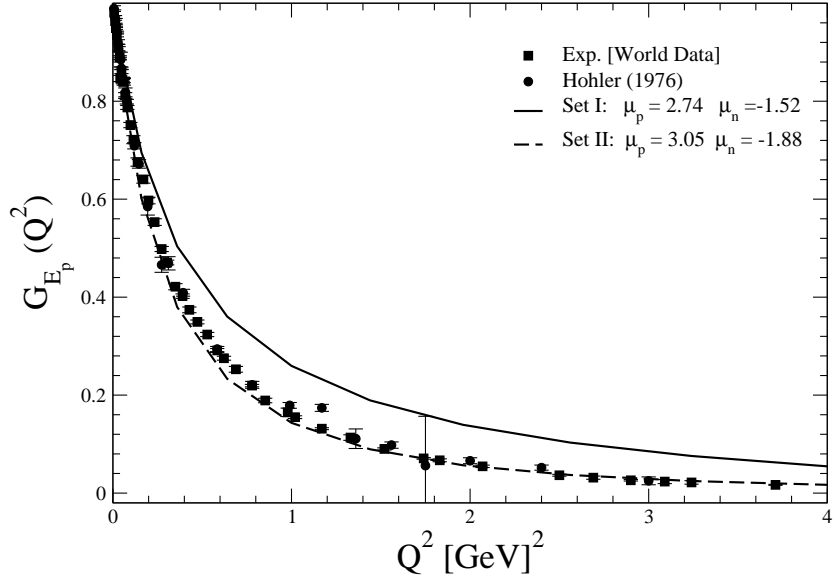}
\\
\vspace*{1.5cm}
\includegraphics[scale=0.44]{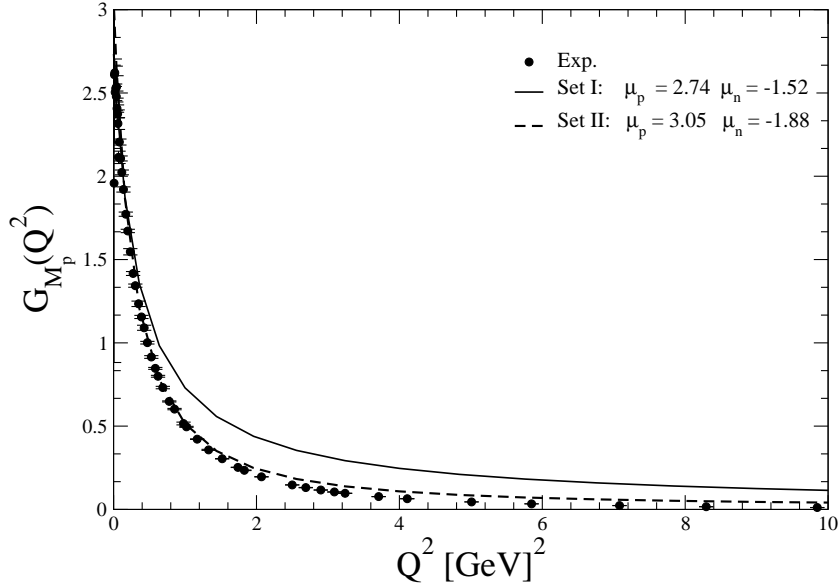}
\caption{Proton electric (upper panel) and magnetic (lower panel) form factors 
calculated by the two-scale model with the two parameter sets, 
set I~(solid line), and set II (dashed line).  
(See also Tab.~\ref{tab1}.)
Experimental data are from 
Refs.~\cite{Jones2000,Gayou2002,Punjabi2005,Ron2007,Puckett2010,Ron2011,Puckett2012,Holer1976,Gao2003}.
}
\label{gepgmp}
\end{figure}

In Fig.~\ref{gepgmp} we show the results for the proton electric (upper panel)  
and magnetic (lower panel) form factors in vacuum, for the two sets of the parameters. 
The experimental data are well described by the set II. 

\begin{figure}[htbp]
\includegraphics[scale=0.5]{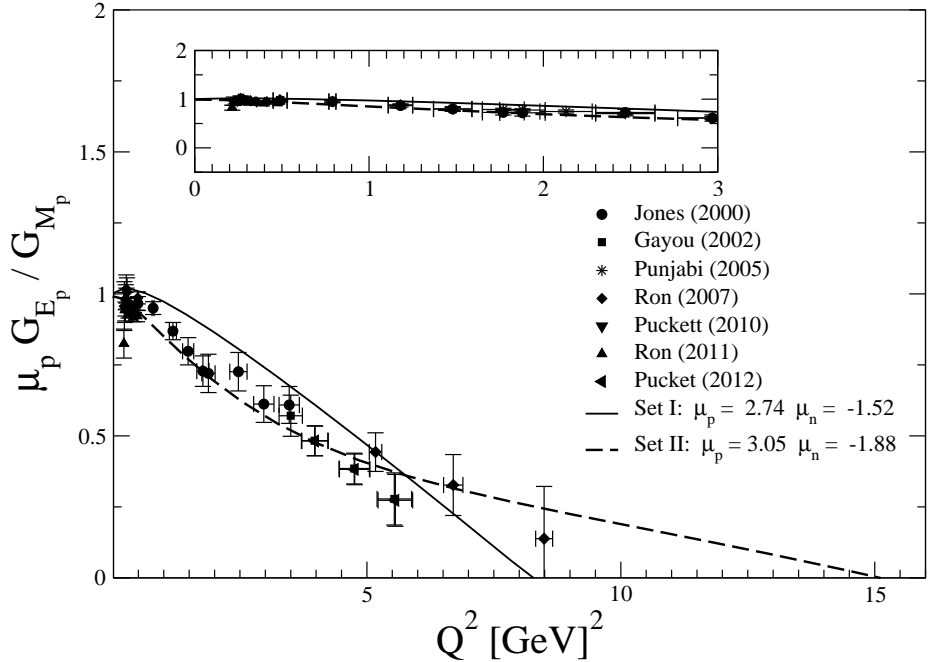}
\caption{
Proton form factor ratio, $\mu_pG_{Ep}(Q^2)/G_{Mp}(Q^2)$, calculated by 
the two-scale model with the two parameter sets, set I and set II.
Experimental data are from Refs.~~\cite{Jones2000,Gayou2002,Punjabi2005,Ron2007,Puckett2010,Ron2011,Puckett2012} 
.}
\label{gepmratio}
\end{figure}

Next, we show in Fig.~\ref{gepmratio} the form factor ratio, 
$\mu_pG_{Ep}(Q^2)/G_{Mp}(Q^2)$, calculated with the two parameter sets, 
the same as those in Fig.~\ref{gepgmp}. 
Reasonable and better agreement with the 
data~\cite{Jones2000,Gayou2002,Punjabi2005,Ron2007,Puckett2010,Ron2011,Puckett2012} is achieved 
by the set II. 
The values of the zero for $G_{Ep}(Q^2)$, $Q_0^2$, are given in Tab.~\ref{tab1} 
for both the set I and set II. The zero of the form factor ratio in vacuum by the set II, 
$Q^2_0 \simeq 15$ GeV$^2$, is one of the main results of this study.
On the other hand, based on the nucleon Bethe-Salpeter amplitude and vector meson dominance model, 
Refs.~\cite{Q02,Q03} report $Q_0^2 \simeq 9$ GeV$^2$. 
In addition, based on the covariant spectator quark model, Ref.~\cite{GilbertoQ0} 
also obtained $Q_0^2 \simeq 9$ GeV$^2$ for $G_{Ep}(Q_0^2)/G_D(Q_0^2) = 0$, with 
$G_D(Q^2) = (1 + Q^2/0.71{\rm GeV}^2)^{-2}$.

\begin{figure}[htbp]
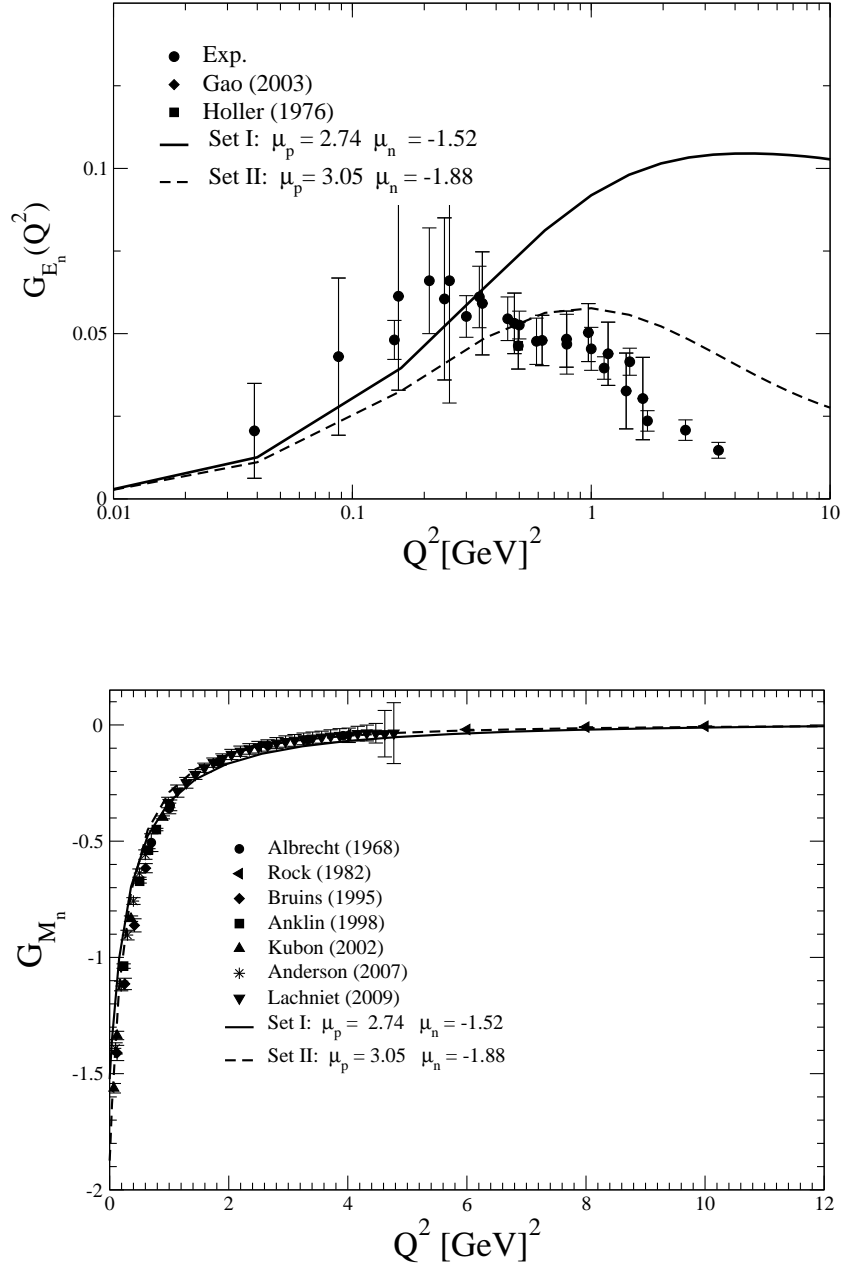

\includegraphics[scale=0.44,angle=0]{genxq2_epja.eps}
\\
\vspace*{1.5cm}
\includegraphics[scale=0.44]{gmnxq2_epja.eps}
\caption{
Neutron electric (upper panel) and magnetic (lower panel) form factors  
calculated by the two-scale model with the two parameter sets.
Experimental data are from~\cite{Holer1976,Gao2003} for~$G_{E_n}$, and   
from~\cite{Albrecht1968,Rock1982,Bruins1995,Anklin1998,Kubon2002,Anderson2007,Lachniet2009} 
for $G_{M_n}$.
}
\label{gengmn}
\end{figure}

We show in Fig.~\ref{gengmn}, the neutron electric (upper panel) and magnetic (lower panel) 
form factors calculated in 
the two-scale model with the two sets of the parameters.   

As one can see, the parameter set II describes better the data due to the strong 
sensitivity to $G_{En}(Q^2)$~\cite{afsbw}.
Note that there is a zero in neutron magnetic form factor for the two-scale nucleon wave functions 
for the both parameter sets. The best fit for the experimental data of neutron magnetic form factor 
is achieved by the set II.

\newpage

\section{Quarks in symmetric nuclear matter: Brief Review}
\label{qmatter}

In order to study the in-medium modifications of the nucleon EM 
form factors, we need a reasonable model of nuclear matter based 
on the quark degrees of freedom as well as the nucleon model in vacuum, 
since our interest is the nucleon internal structure change in a nuclear medium. 

As for the model of nuclear matter, we use the quark-meson coupling (QMC) model, 
which has been successfully applied for the studies of light-quark flavor as well as 
strange and charm hadron properties in a nuclear medium, 
and finite (hyper)nuclei~\cite{QMCreview,QMCfinite}. 
(The QMC model in early stage was also applied for finite nuclei 
in Ref.~\cite{BlundenFN}. For more references on the ``QMC'' model from 
various other groups, readers are asked to consult Ref.~\cite{QMCreview}.)  
This model was already used for the study of the in-medium pion 
properties~\cite{pimedium1,pimedium2,pimedium3,pimedium4} 
combined with a light-front constituent 
quark model. Therefore, we can extend the study of pion properties in medium 
for the in-medium nucleon EM form factors in a similar manner.
We first review the quark model description of nuclear matter via the QMC model, 
and present the results for nuclear saturation properties  
as well as the in-medium constituent up and down quark properties,  
the same inputs used in the study of pion properties 
in medium~\cite{pimedium1,pimedium2,pimedium3,pimedium4}.

\subsection{Quark model of nuclear matter: quark-meson coupling (QMC) model}
\label{QMC}

The QMC model was introduced by Guichon~\cite{Guichon} in 1988  
using the MIT bag model, and also by Frederico {\it et al\/}. 
in 1989~\cite{Frederico} using a confining harmonic potential,
to describe the properties of nuclear matter based on the quark degrees of freedom. 
(See Ref.~\cite{QMCreview} for other references.) 
In this study we use the model of Guichon with the MIT bag model. 
The model has been successfully applied for various studies of finite
(hyper)nuclei~\cite{QMCfinite} as well as the hadron properties 
in a nuclear medium~\cite{QMCreview}.
In the model the medium effects arise from the self-consistent direct coupling of
the isoscalar-Lorentz-scalar ($\sigma$), isoscalar-Lorentz-vector ($\omega$) 
and isovector-Lorentz-vector ($\rho$) meson mean fields  
to the confined light-flavor $u$ and $d$ valence quarks --- rather than to the nucleons.
As a result the internal structure of the bound nucleon is modified by the
surrounding nuclear medium with respect to the free nucleon.

\begin{table}[tb]
\begin{center}
\caption{Coupling constants, the parameter $Z_N$, bag constant $B$ (in $B^{1/4}$),
and calculated properties for symmetric nuclear matter
at normal nuclear matter density $\rho_0 = 0.15$ fm$^{-3}$,
for $m_q = 5$ and $220$ MeV (the latter values is used in this study and Refs.~\cite{pimedium1,pimedium2,pimedium3,pimedium4}). 
The effective nucleon mass, $m_N^*$, and the nuclear
incompressibility, $K$, are quoted in MeV (the free nucleon bag radius input is $R_N = 0.8$ fm,
the standard value in the QMC model~\cite{QMCreview}). }
\label{Tab:QMC}
\bigskip
\begin{tabular}{c|cccccc}
\hline
$m_q$(MeV)&$g_{\sigma}^2/4\pi$&$g_{\omega}^2/4\pi$
&$m_N^*$ &$K$ & $Z_N$ & $B^{1/4}$(MeV)\\
\hline
 5   &5.39 &5.30 &754.6 &279.3 &3.295 &170 \\
 220 &6.40 &7.57 &698.6 &320.9 &4.327 &148 \\
\hline
\end{tabular}
\end{center}
\end{table}

The effective Lagrangian density of the QMC model for a uniform, spin-saturated,
and isospin-symmetric infinite nuclear matter 
at the hadronic level is given by~\cite{Guichon,QMCfinite,QMCreview},
\begin{equation}
{\cal L} = {\bar \psi} [i\gamma \cdot \partial -m_N^*({\hat \sigma}) -g_\omega {\hat \omega}^\mu \gamma_\mu ] \psi
+ {\cal L}_\textrm{meson} ,
\label{lag1}
\end{equation}
where $\psi$, ${\hat \sigma}$ and ${\hat \omega}$ are respectively the nucleon,
Lorentz-scalar-isoscalar $\sigma$, and Lorentz-vector-isoscalar $\omega$ field operators, with
\begin{equation}
m_N^*({\hat \sigma}) \equiv m_N - g_\sigma({\hat \sigma}) {\hat \sigma} ,
\label{efnmas}
\end{equation}
which defines the $\sigma$-field dependent coupling constant,
$g_\sigma({\hat \sigma})$, while $g_\omega$ is the nucleon-$\omega$ coupling
constant. All the important effective nuclear many-body dynamics including 3-body
nucleon force modeled at the quark level, can be regarded as 
effectively condensed in $g_\sigma({\hat \sigma})$.
Solving the Dirac equations for the up and down quarks in the nuclear medium with
the same mean fields (mean values) $\sigma$ and $\omega$ which act on 
the bound nucleon self-consistently based on Eq.~(\ref{lag1}), 
we obtain the effective $\sigma$-dependent coupling $g_\sigma(\sigma)$
at the nucleon level~\cite{Guichon,QMCfinite,QMCreview}. 
The free meson Lagrangian density is given by,
\begin{equation}
{\cal L}_\mathrm{meson} = \frac{1}{2} (\partial_\mu {\hat \sigma} 
\partial^\mu {\hat \sigma} - m_\sigma^2 {\hat \sigma}^2)
- \frac{1}{2} \partial_\mu {\hat \omega}_\nu (\partial^\mu {\hat \omega}^\nu - \partial^\nu {\hat \omega}^\mu)
+ \frac{1}{2} m_\omega^2 {\hat \omega}^\mu {\hat \omega}_\mu \ ,
\label{mlag1}
\end{equation}
where we have ignored the isospin-dependent Lorentz-vector-isovector $\rho$-meson field,
since we consider isospin-symmetric nuclear matter within the Hartree mean-field approximation. 
In this case the mean value of the $\rho$-meson field becomes zero and there is
no need to consider its possible contributions due to the $\rho$-Fock (exchange) terms.

In the following we work in the nuclear matter rest frame.
For symmetric nuclear matter in the Hartree mean-field approximation, 
the nucleon Fermi momentum $k_F$ (baryon density $\rho$) 
and the scalar density ($\rho_s$) associated with
the $\sigma$-mean field can be related as,
\begin{eqnarray}
\rho &=& \frac{4}{(2\pi)^3}\int d^3k\ \theta (k_F - |\vec{k}|)
= \frac{2 k_F^3}{3\pi^2},
\label{rhoB}\\
\rho_s &=& \frac{4}{(2\pi)^3}\int d^3k \ \theta (k_F - |\vec{k}|)
\frac{m_N^*(\sigma)}{\sqrt{m_N^{* 2}(\sigma)+\vec{k}^2}},
\label{rhos}
\end{eqnarray}
where $m_N^*(\sigma)$ is the constant value of the effective nucleon
mass at a given density, and is calculated in the standard QMC model~\cite{Guichon,QMCfinite,QMCreview}.
The Dirac equations for the up ($u$) and down ($d$) 
quarks in symmetric nuclear matter are solved
self-consistently with the same $\sigma$ and $\omega$ mean-field potentials 
acting for the nucleon.
We restrict ourselves hereafter the flavor SU(2), the $u$ and $d$ quark sector 
(as well as for the proton and neutron).
The Dirac equations for the quarks and antiquarks
($q = u$ or $d$, quarks)
in the bag of hadron $h$ in nuclear matter at the position
$x=(t,\vec{r})$ ($|\vec{r}| \le$ bag radius) are given by~\cite{QMCreview},
\begin{eqnarray}
\left[ i \gamma \cdot \partial_x -
(m_q - V^q_\sigma)
\mp \gamma^0
\left( V^q_\omega +
\frac{1}{2} V^q_\rho
\right) \right]
\left( \begin{array}{c} \psi_u(x)  \\
\psi_{\bar{u}}(x) \\ \end{array} \right) &=& 0,
\label{diracu}\\
\left[ i \gamma \cdot \partial_x -
(m_q - V^q_\sigma)
\mp \gamma^0
\left( V^q_\omega -
\frac{1}{2} V^q_\rho
\right) \right]
\left( \begin{array}{c} \psi_d(x)  \\
\psi_{\bar{d}}(x) \\ \end{array} \right) &=& 0,
\label{diracd}
%
\end{eqnarray}
where we have neglected the Coulomb force as usual, since the nuclear matter
properties are due to the strong interaction, and we assume SU(2) symmetry for 
the light-flavor $u$ and $s$ quarks,
$m_q = m_u = m_d$, and define $m^*_q \equiv m_q - V^q_\sigma = m^*_u = m^*_d$.
In symmetric nuclear matter, the isospin dependent $\rho$-meson mean field
in Hartree approximation yields $V^q_\rho = 0$ in Eqs.~(\ref{diracu})
and~(\ref{diracd}), as mentioned already, so we ignore it hereafter.
The constant mean-field potentials in (symmetric) nuclear matter are defined by,
$V^q_\sigma \equiv g^q_\sigma \sigma = g^q_\sigma <\sigma>$ and
$V^q_\omega \equiv g^q_\omega \omega = g^q_\omega\, \delta^{\mu,0} <\omega^\mu>$,
with $g^q_\sigma$ and $g^q_\omega$ being the corresponding quark-meson coupling constants,
and the quantities inside the brackets stand for taking expectation values
by the nuclear matter ground state~\cite{QMCreview}.
Note that, since the velocity averages to zero in the rest frame of nuclear matter,
the mean vector source due to the quark fields as well,
$<\bar{\psi_q} \vec{\gamma} \psi_q> = 0$.
Thus we may just keep the term proportional to $\gamma^0$ in Eqs.~(\ref{diracu})
and~(\ref{diracd}).

The normalized, static solution for the ground state quarks or antiquarks
with flavor $f$ in the hadron $h$ composed of $u$ and $d$ quarks, may be written,
$\psi_f (x) = N_f e^{- i \epsilon_f t / R_h^*}
\psi_f (\vec{r})$,
where $N_f$ and $\psi_f(\vec{r})$
are the normalization factor and
corresponding spin and spatial part of the wave function.
The bag radius in medium for a hadron $h$, $R_h^*$,
is determined through the
stability condition for the mass of the hadron against the
variation of the bag radius~\cite{QMCreview}.
The eigenenergies in units of $1/R_h^*$ are given by,
\bge
\left( \begin{array}{c}
\epsilon_u \\
\epsilon_{\bar{u}}
\end{array} \right)
= \Omega_q^* \pm R_h^* \left(
V^q_\omega
+ \frac{1}{2} V^q_\rho \right),\,\,
\left( \begin{array}{c} \epsilon_d \\
\epsilon_{\bar{d}}
\end{array} \right)
= \Omega_q^* \pm R_h^* \left(
V^q_\omega
- \frac{1}{2} V^q_\rho \right).
%
\label{energy}
\ene

\begin{figure}[tb]
\begin{center}
\includegraphics[scale=0.5]{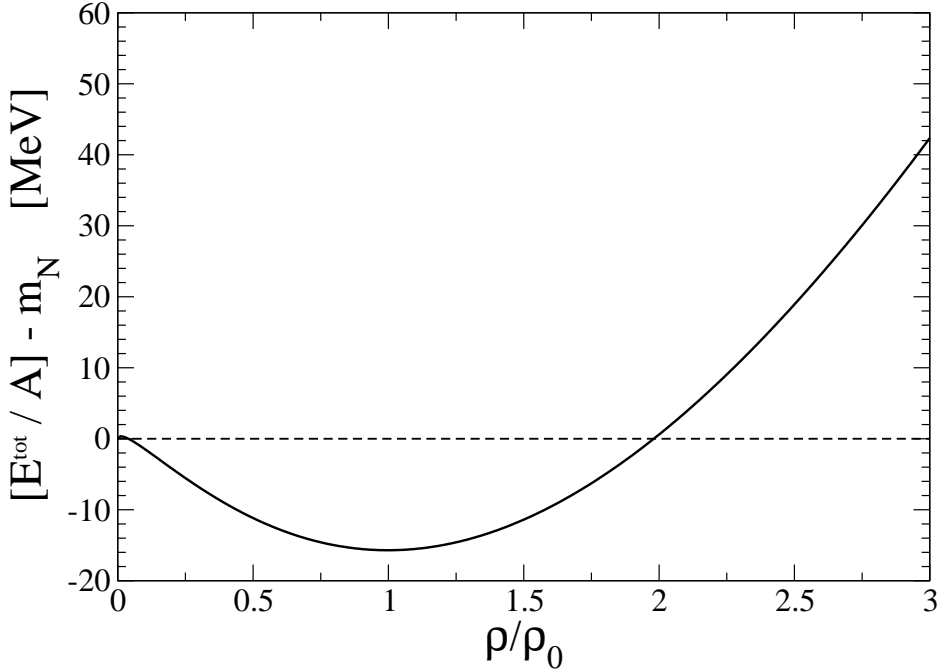}
\caption{Negative of the binding energy per nucleon ($E^\mathrm{tot}/A - m_N$) for symmetric
nuclear matter calculated with the vacuum up and down quark mass, $m_q = 220$ MeV, 
taken from Ref.~\cite{pimedium1,pimedium2,pimedium3,pimedium4}.
At the saturation point $\rho_0 = 0.15$ fm$^{-3}$, the value is fitted to $-15.7$~MeV.
(See Ref.~\cite{QMCreview} for the $m_q = 5$ MeV case, denoted in there as QMC-I.)
\label{Fig:Eden}
}
\end{center}
\end{figure}

\begin{figure}[tb]
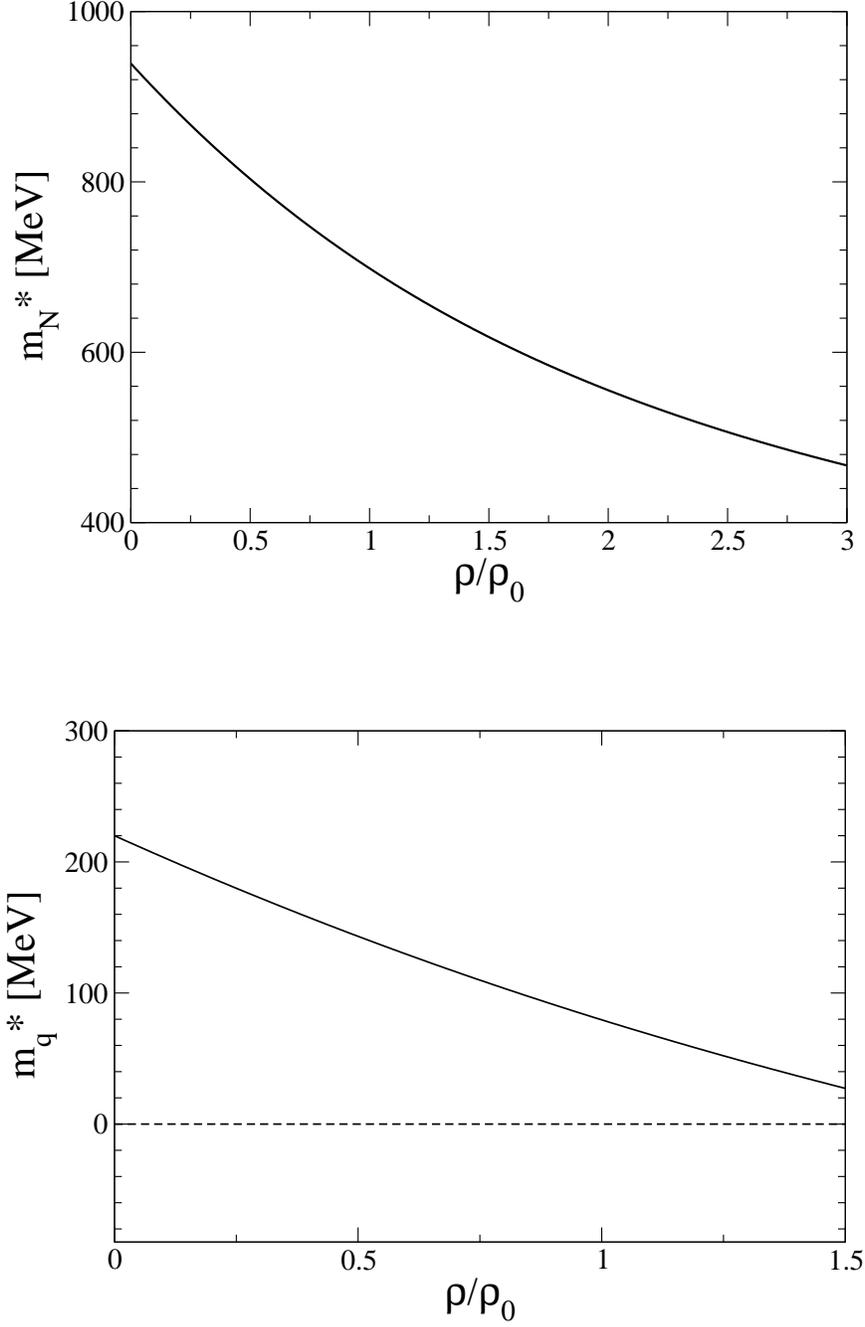

\begin{center}
\includegraphics[scale=0.45]{mNstar.eps}
\\
\vspace*{1.5cm}
\includegraphics[scale=0.45]{mqstar2.eps}
\caption{Nucleon and constituent quark effective masses, respectively $m^*_N$ (upper panel), 
and $m^*_q$ (lower panel) where $m^*_q \equiv m^*_u = m^*_d$, in symmetric nuclear 
matter taken from Refs.~\cite{pimedium1,pimedium2,pimedium3,pimedium4}. 
See also caption of Fig.~\ref{Fig:Eden}.
\label{Fig:mNmqstar}
}
\end{center}
\end{figure}

The hadron masses
in a nuclear medium $m^*_h$ (free mass $m_h$),
are calculated by
\begin{eqnarray}
m_h^* &=& \sum_{j=q,\bar{q}}
\frac{ n_j\Omega_j^* - z_h}{R_h^*}
+ \frac{4}{3}\pi R_h^{* 3} B,\quad
\left. \frac{\partial m_h^*}
{\partial R_h}\right|_{R_h = R_h^*} = 0,
\label{hmass}
\end{eqnarray}
where $\Omega_q^*=\Omega_{\bar{q}}^*
=[x_q^2 + (R_h^* m_q^*)^2]^{1/2}$, with
$m_q^*=m_q{-}g^q_\sigma \sigma$,
and $x_{q}$ being the lowest bag eigenfrequencies.
$n_q (n_{\qbar})$ 
is the quark (antiquark)
numbers for the quark flavors $q$.
The MIT bag quantities, $z_h$, $B$, $x_q$,
and $m_q$ are the parameters for the sum of the c.m. and gluon
fluctuation effects, bag constant, lowest eigenvalues for the quarks
$q$, and the corresponding current quark masses.
$z_N$ and $B$ ($z_h$) are fixed by fitting the nucleon
(the hadron) mass in free space. (See Tab.~\ref{Tab:QMC} for the nucleon case.)

For the nucleon $h=N$ in the above, the lowest, positive energy bag eigenfunction is given by
\begin{equation}
q(t,\vec{r}) = \frac{\cal N}{\sqrt{4\pi}}
e^{-i\epsilon_qt/R^*_N}\left( \begin{array}{c}
j_0(xr/R^*_N) \\
i\beta_q {\vec \sigma}\cdot{\hat r} j_1(xr/R^*_N)
\end{array} \right) \theta(R^*_N-r) \chi_m ,
\label{lowest}
\end{equation}
with $r=|\vec{r}|$ and $\chi_m$ the spin function and
\begin{eqnarray}
\Omega_q^* &=& \sqrt{x^2+(m_q^\ast R^*_N)^2}, \ \
\beta_q = \sqrt{\frac{\Omega_q^* - m_q^\ast R^*_N}{\Omega_q^* + m_q^\ast R^*_N}} ,
\label{omega0} \\
{\cal N}^{-2} &=& 2R_N^{*3} j_0^2(x)[\Omega_q^*(\Omega_q^*-1) + m_q^\ast R^*_N/2]/x^2  ,
\label{norm}
\end{eqnarray}
where $x$ is the eigenvalue for the lowest mode,
which satisfies the boundary condition at the bag surface,
$j_0(x) = \beta_q j_1(x)$ with $j_{0,1}$ are the spherical Bessel functions.

The same meson mean fields $\sigma$ and $\omega$ for the quarks and nucleons satisfy
the following equations at the nucleon level self-consistently:
\begin{eqnarray}
{\omega}&=&\frac{g_\omega \rho}{m_\omega^2},
\label{omgf}\\
{\sigma}&=&\frac{g_\sigma }{m_\sigma^2}C_N({\sigma})
\frac{4}{(2\pi)^3}\int d^3 k \ \theta (k_F - |\vec{k}|)
\frac{m_N^*(\sigma)}{\sqrt{m_N^{* 2}(\sigma)+\vec{k}^2}} 
= \frac{g_\sigma }{m_\sigma^2}C_N({\sigma}) \rho_s,
\label{sigf}\\
C_N(\sigma) &=& \frac{-1}{g_\sigma(\sigma=0)}
\left[ \frac{\partial m^*_N(\sigma)}{\partial\sigma} \right],
\label{CN}
\end{eqnarray}
where $C_N(\sigma)$ is the constant value of the scalar density ratio~\cite{Guichon,QMCfinite,QMCreview}.
Because of the underlying quark structure of the nucleon used to calculate
$M^*_N(\sigma)$ in the nuclear medium (see Eq.~(\ref{hmass}) with $h=N$),
$C_N(\sigma)$ gets nonlinear $\sigma$-dependence,
whereas the usual point-like nucleon-based model yields unity, $C_N(\sigma) = 1$.
It is this $C_N(\sigma)$ or $g_\sigma (\sigma)$ that gives a novel saturation mechanism
in the QMC model, and contains the important dynamics which originates from the quark structure
of the nucleon. Without an explicit introduction of the nonlinear
couplings of the meson fields in the Lagrangian density at the nucleon and meson level,
the standard QMC model yields the nuclear incompressibility of $K \simeq 280$~MeV with 
$m_q=5$ MeV, which is in contrast to a naive version of quantum hadrodynamics (QHD)~\cite{QHD}
(the point-like nucleon model of nuclear matter),
results in the much larger value, $K \simeq 500$~MeV;
the empirically extracted value falls in the range $K = 200 - 300$ MeV.
(See Ref.~\cite{Stone} for the updated discussions on the incompressibility.)

Once the self-consistency equation for the ${\sigma}$, Eq.~(\ref{sigf}), 
has been solved, one can evaluate the total energy per nucleon:
\begin{equation}
E^\mathrm{tot}/A=\frac{4}{(2\pi)^3 \rho}\int d^3 k \
\theta (k_F - |\vec{k}|) \sqrt{m_N^{* 2}(\sigma)+
\vec{k}^2}+\frac{m_\sigma^2 {\sigma}^2}{2 \rho}+
\frac{g_\omega^2\rho}{2m_\omega^2} .
\label{toten}
\end{equation}
We then determine the coupling constants, $g_{\sigma}$ and $g_{\omega}$, so as
to fit the binding energy of 15.7~MeV at the saturation density $\rho_0$ = 0.15 fm$^{-3}$
($k_F^0$ = 1.305 fm$^{-1}$) for symmetric nuclear matter.

In the study of pion properties in medium~\cite{pimedium1,pimedium2,pimedium3,pimedium4} 
based on a light-front constituent quark model, the vacuum value of the $u$ and $d$ quark 
constituent mass, $m_q = 220$~MeV was used and could reproduce well 
the EM form factor and decay constant in vacuum~\cite{Pacheco1}.

To be consistent and encouraged by the studies for the pion properties 
in a nuclear medium~\cite{pimedium1,pimedium2,pimedium3,pimedium4}, 
we build the nuclear matter with the same $u$ and $d$ constituent quark mass in vacuum. 
The corresponding coupling constants and some results for symmetric nuclear matter 
at the saturation density calculated with $m_q = 220$ MeV and the standard values of 
$m_{\sigma}=550$ MeV and $m_{\omega}=783$~MeV, are listed in Tab.~\ref{Tab:QMC}.
For comparison, we also give the corresponding quantities calculated in the standard QMC
model with a vacuum quark mass of $m_q = 5$~MeV (see Ref.~\cite{QMCreview} for details).
Thus, we have obtained the in-medium properties of the $u$ and $d$ constituent quarks in symmetric
nuclear matter with the vacuum mass of $m_q = 220$~MeV. Namely, we obtain 
the density dependence of the effective mass (scalar potential) and vector potential.
Using the obtained in-medium inputs, we study the nucleon 
EM form factors in symmetric nuclear matter.

In Figs.~\ref{Fig:Eden} and~\ref{Fig:mNmqstar}, we respectively show the results for
the negative of the binding energy per nucleon ($E^\mathrm{tot}/A - m_N$),
effective mass of the nucleon, $m_N^*$,
and effective mass of the constituent up and down quarks, 
$m_q^*$, in symmetric nuclear matter.

As one can expect from the values of the incompressibility,~$K = (279.3, 320.9)$~MeV 
for~$m_q~(5,220)$~MeV, in Tab.~\ref{Tab:QMC}, 
the result for $E/A - m_N$ with $m_q = 220$ MeV shown in Fig.~\ref{Fig:Eden} varies slightly 
faster than that for the case of $m_q = 5$~MeV~\cite{QMCreview} as increasing nuclear matter density. 
As for the effective nucleon mass shown in Fig.~\ref{Fig:mNmqstar} with $m_q = 220$~MeV, 
it also decreases faster than that for $m_q = 5$ MeV~\cite{QMCreview}   
as increasing nuclear matter density.

In next section we study the nucleon EM form factors 
in a nuclear medium using the in-medium constituent quark properties 
obtained so far.

\section{Nucleon electromagnetic form factors in medium}
\label{mediumffs}

In this section we present our main results, the nucleon EM form factors 
in symmetric nuclear matter, $G^*_{Ep}(Q^2), G^*_{Mp}(Q^2), G^*_{En}(Q^2), G^*_{Mn}(Q^2)$, 
the ratio, $\mu_p G^*_{Ep}(Q^2)/G^*_{Mp}(Q^2)$, and the double ratio, 
$R_p \equiv [G^*_{Ep}(Q^2)/G^*_{Mp}(Q^2)]/[G_{Ep}(Q^2)/G_{Mp}(Q^2)]$, 
and the corresponding double ratio for neutron, $R_n$.
Our interests in this section are, in-medium effects on the zero of 
$\mu_p G^*_{Ep}(Q^2)/G^*_{Mp}(Q^2)$, 
and the comparison with the JLab data for   
$[G^{^4\rm He}_{Ep}(Q^2)/G^{^4\rm He}_{Mp}(Q^2)]/[G^{^1\rm H}_{Ep}(Q^2)/G^{^1\rm H}_{Mp}(Q^2)]$, 
as well as the corresponding neutron EM form factor double ratio $R_n$ 
in symmetric nuclear matter.

Before presenting the results for the in-medium nucleon EM 
form factors, we briefly explain how the nucleon is treated 
in medium within the present model. The description of the nucleon in vacuum, 
and the calculation of the relevant microscopic EM matrix elements 
are explained in Appendix~\ref{appendix}. 
The starting point is the relativistic invariant effective Lagrangian density in vacuum 
Eq.~(\ref{lag}). Then the calculation in vacuum is made by the light-front 
projection with the {\it relativistic quark-spin coupling model}, 
a light-front constituent quark model. 
Although it would be idealistic to construct the nuclear matter 
also within the same model with the nucleon substructure, 
it would be a very difficult task to achieve properly with being guaranteed 
by phenomenological success. 
In particular, this is true when we try to describe the nuclear matter 
based on the quark degrees of freedom within the same model.
In fact, light-front based nuclear mean field theory~\cite{LFNMean} 
and some applications exist~\cite{LFEMC},  
but we would like to take a practical 
manner to adapt the in-medium inputs necessary from the 
quark-based successful model, the QMC model~\cite{QMCreview,QMCfinite}, 
which has already been explained in Sec.~\ref{qmatter}. 

What characterize the in-medium properties of nucleon and $u$ and $d$ quarks in medium 
are, the Lorentz-scalar and Lorentz-vector mean field potentials felt by 
the nucleon and the $u$ and $d$ quarks in medium, consistently obtained with the 
nuclear matter saturation properties. For this purpose, we rely on the 
quark-based successful model, the QMC model.  
Then, based on the explanations given in Appendix~\ref{appendix}, 
the in-medium treatment is made as follows. 
The momentum of the light quark $j \hspace{1ex} (j=1,2,3)$ in the nucleon,    
$k_j^\mu$, is replaced by $k_j^{*\mu} = (k_j^0 + V^q_\omega, \vec{k}_j)$,  
where $V^q_\omega$ is the vector mean field potential  
felt by a light-flavor quark in symmetric nuclear matter. 
(Space component of the momentum is not modified in the present 
case of Hartree mean field approximation.)
Correspondingly, the in-medium light-front momentum is defined by  
$k^{*+}_j = k^{*0}_j + k^3_j$.
Since the nucleon consists of three light quarks, 
its momentum in free space $p^\mu$ is replaced by   
$p^{* \mu} = (p^0 + 3 V^q_\omega, \vec{p}) 
= (\sqrt{m_N^{*2} + \vec{p}^{\,2}} + 3 V^q_\omega, \vec{p})$, and thus 
the corresponding in-medium light-front plus-momentum becomes 
$p^{*+} = p^{*0} + p^3 = (\sqrt{m_N^{*2} + \vec{p}^{\,2}} + 3 V^q_\omega) + p^3$. 
Furthermore, the quark and nucleon masses in vacuum $m_q$ and $m_N$ are respectively 
replaced by $m_q \to m^*_q$ and $m_N \to m^*_N$, whenever they appear 
in the expressions in vacuum.
Note that, for the Dirac particle spinor in medium with its three-momentum $\vec{k}$, 
the energy $E_N^*(\vec{k}) = \sqrt{m_N^{* 2} + \vec{k}^{\,2}}$ is used 
without the vector potential $3 V^q_\omega$~\cite{QHDHF}.
These in-medium inputs $m_q^*, V^q_\omega$ and $m_N^*$, are calculated 
by the QMC model for a given nuclear matter density as explained above.
(See also Sec.~\ref{qmatter}.) 

In the following we present results for the in-medium nucleon EM form factors.
We note that the nuclear matter densities $\rho=0.3\rho_0$ and $0.4\rho_0$ 
studied in this section (except for the set II in Fig.~\ref{figratio}),    
are chosen so that to give a trend of medium effects, 
based on a rough estimate made for the proton EM double ratio 
by the set I to be shown later in Fig.~\ref{figratio}.
First, we give in Tab.~\ref{tab_medium} some static properties 
of nucleon in medium together with those in vacuum.
\begin{table}[htbp]
\caption{Nucleon EM static observables and the zero of
$G_{Ep}(Q^{2 *}_0) = 0$, $Q^{2 *}_0$, for the two-scale models 
with the two sets of the parameters, set I and set II for 
densities $\rho/\rho_0 = 0.0, 0.3$ and $0.4$.  
(See also caption of Tab.~\ref{tab1}.)
\label{tab_medium}
}
\begin{center}
\begin{tabular}{|l|c|c|c|c|c|c|c|}
\hline 
\hline
 & $\rho/\rho_{0}$ &$\mu_p^*$ &~$r_p^*$~(fm) & $\mu_n^*$&
$r^{2 *}_n$~(fm)$^2$ & $Q_0^{2 *}$~GeV$^2$ \\
\hline
Set I & 0.0   & 2.74   &~0.80  & -1.52 & -0.07 & 8.27  \\
      & 0.3   & 2.91   &~0.95  & -1.68 & -0.13 & 7.12  \\
      & 0.4   & 2.96   &~1.00  & -1.72 & -0.15 & 6.78  \\
\hline
Set II & 0.0   & 3.05 &~0.94  & -1.88 & -0.06 & 15.12  \\
       & 0.3   & 3.23 &~1.08  & -2.05 & -0.11 & 6.34  \\
       & 0.4   & 3.29 &~1.18  & -2.11 & -0.12 & 5.10  \\
\hline
\hline 
\end{tabular}
\end{center}
\end{table}

One can notice that the magnetic moments of proton and neutron in medium 
are enhanced as nuclear matter density increases for the both parameters sets I and II.
So do the in-medium proton root-mean-square charge radius $r^*_p$  
and neutron mean-square charge radius $r_n^{* 2}$. 
These features are in agreement with those found in Ref.~\cite{Ding} 
studied in the QMC model. 
One of the very interesting quantities is $Q_0^{2 *}$, the value 
of crossing zero, namely the value satisfying $G^*_{Ep}(Q_0^{2 *}) = 0$.
The values of $Q_0^{2 *}$ decrease as nuclear matter density increases   
for the both parameter sets I and II. 
The corresponding figure will be shown in Fig.~\ref{mupgepgmp_med}.

\begin{figure}[htbp]
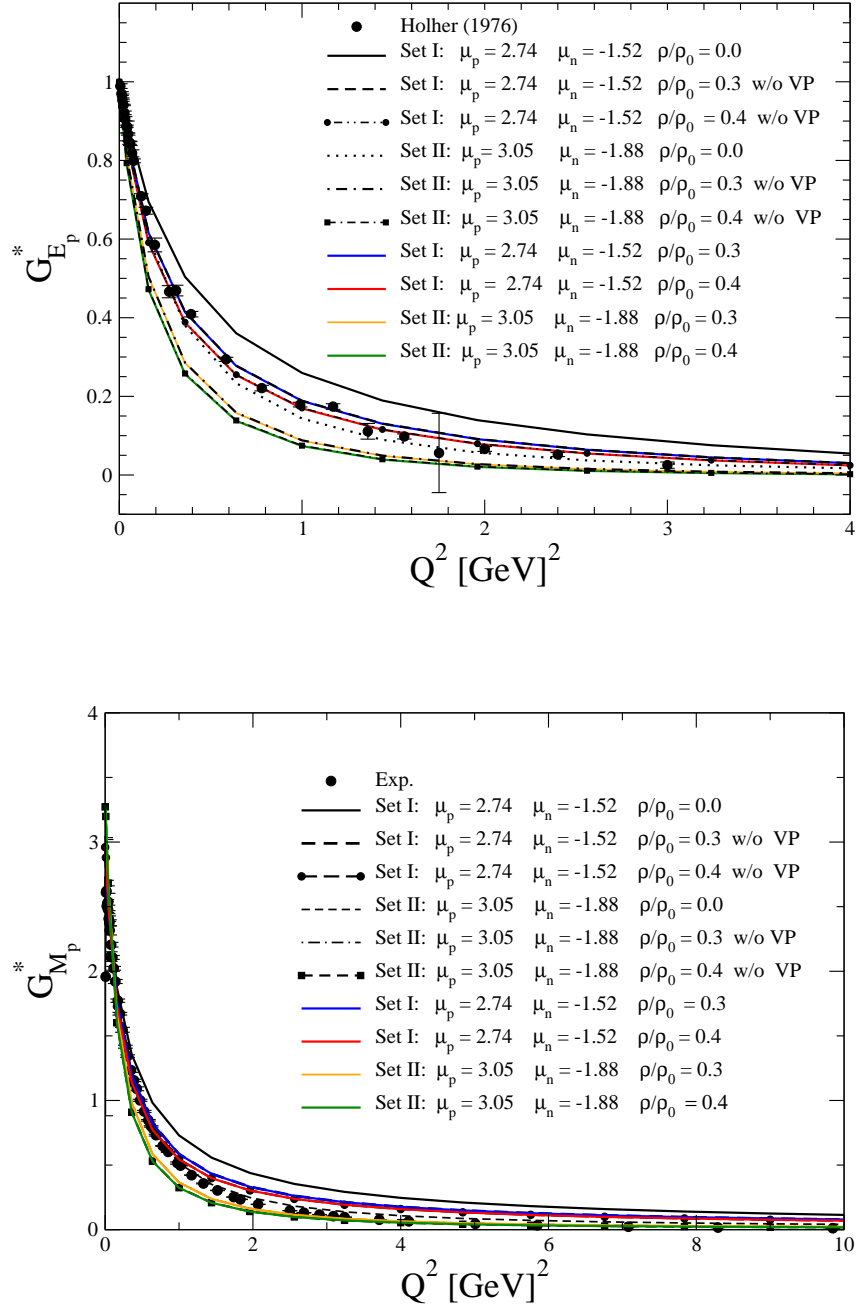

\includegraphics[scale=0.45,angle=0]{gep_medium_new.eps}
\\
\vspace*{1.5cm}
\includegraphics[scale=0.455,angle=0]{gmp_medium_new.eps}
\caption{
Proton electric $G^*_{Ep}(Q^2)$ (upper panel),  
and magnetic form $G^*_{Mp}(Q^2)$ (lower panel) form factors  
calculated in the two-scale model with the two parameter sets, set I and set II,  
for nuclear matter densities $\rho = 0.3\rho_0$ and $0.4\rho_0$ with $\rho_0=0.15$ fm$^{-3}$, 
together with those in vacuum. (See also Tab.~\ref{tab1}.) 
In figure, ``w/o VP'' stands for the result calculated without the vector potential.  
Experimental world data are 
from~\cite{Arrington2007,MainzR1,Passchier99,Eden94,JlabR2,Riordan10,Schiavilla01}.
}  
\label{gepgmp_medium}
\end{figure}

\begin{figure}[htbp]
\vspace*{1.2cm}
\includegraphics[scale=0.5,angle=0]{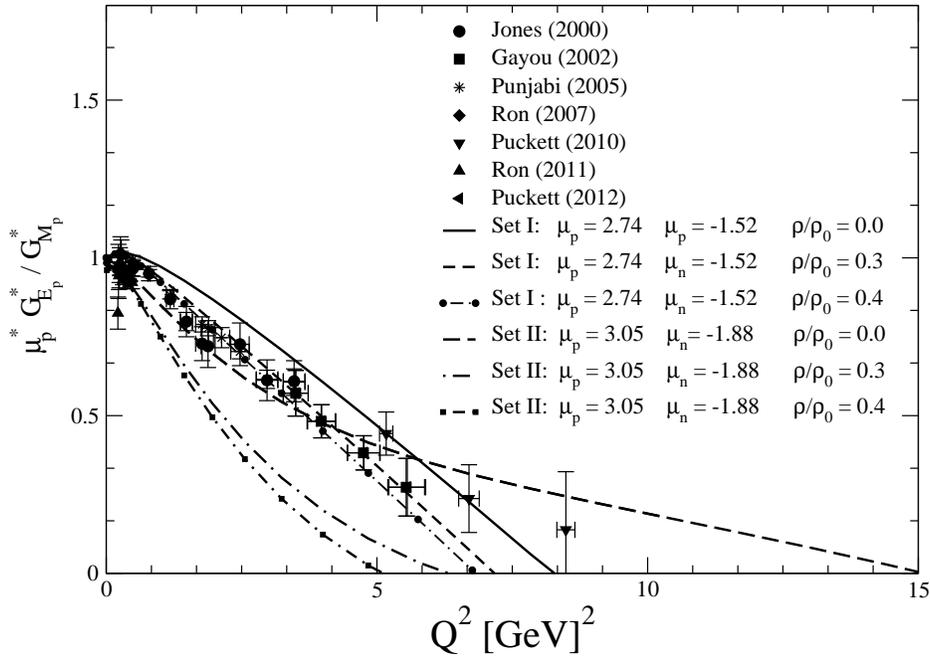}
\caption{
$\mu^*_p G^*_{Ep}(Q^2)/G^*_{Mp}(Q^2)$ calculated by the two-scale model with the two 
parameter sets, for the nuclear matter densities $\rho = 0.3\rho_0$ and $0.4\rho_0$.
} 
\label{mupgepgmp_med}
\end{figure}

Next, we show in Fig.~\ref{gepgmp_medium} in-medium proton electric 
$G^*_{Ep}(Q^2)$ (upper panel) and magnetic $G^*_{Mp}$ (lower panel) form factors 
in the two-scale model with the two parameter sets, set I and set II,  
for nuclear matter densities of $\rho=0.3\rho_0$ and $0.4\rho_0$ ($\rho_0=0.15$ fm$^{-3}$), 
together with those in vacuum to make easier to see the medium effects.
For $G^*_{Ep}(Q^2)$, the falloff becomes faster as increasing nuclear matter density than   
that in vacuum. This means that the proton mean-square charge radius 
increases in symmetric nuclear matter.
The fast falloff of the electric form factor was also 
found in Refs.~\cite{Ding,Gilberto}.
From this behavior, we can expect that the zero of $G^*_{Ep}(Q^2)$ in medium 
shifts to a smaller $Q^2$ value.

As for $G^*_{Mp}(Q^2)$, the in-medium proton magnetic moment $\mu^*_p = G^*_{Mp}(0)$ is 
enhanced than that in vacuum as increasing nuclear matter density. 
This enhancement is also, observed in Refs.~\cite{Ding,Gilberto}.
However, as increasing $Q^2$, the falloff of the in medium one, $G^*_{Mp}(Q^2)$,  
becomes faster than that in vacuum.

To understand better the in-medium effect on the nucleon EM form factors, we show 
the results without the vector potential, namely, the results only included the 
effects of the nucleon and quark mass shits in medium for 
$G^*_{Ep}(Q^2)$ (upper panel) and $G^*_{Mp}(Q^2)$ (lower panel). 
They are denoted by ``w/o VP'' in Fig.~\ref{gepgmp_medium} for $\rho/\rho_0 = 0.3$ and $0.4$. 
One can see that the effect of the vector potential is very small 
for both parameter sets I and II. Typically the effect is a few percent at most 
for the corresponding density, and cannot be distinguished well 
from the full result with the vector potential.
This feature is also reflected from the densities treated here are 
relatively small, and thus the effect of the vector potential becomes small.
Nevertheless, we can see the effect of the mass shifts is larger than that 
of the vector potential in the present model.
We have also studied the effect of the vector potential for all the other 
EM form factors, and confirmed that the effect is small. 
Thus, we will not show the other EM form factor results calculated 
without the vector potential.

Based on the results shown in Fig.~\ref{gepgmp_medium}, 
we show in Fig.~\ref{mupgepgmp_med} the result for 
$\mu^*_p G^*_{Ep}(Q^2) / G^*_{Mp}(Q^2)$, in symmetric nuclear matter as well as in vacuum.
\begin{figure}[htbp]
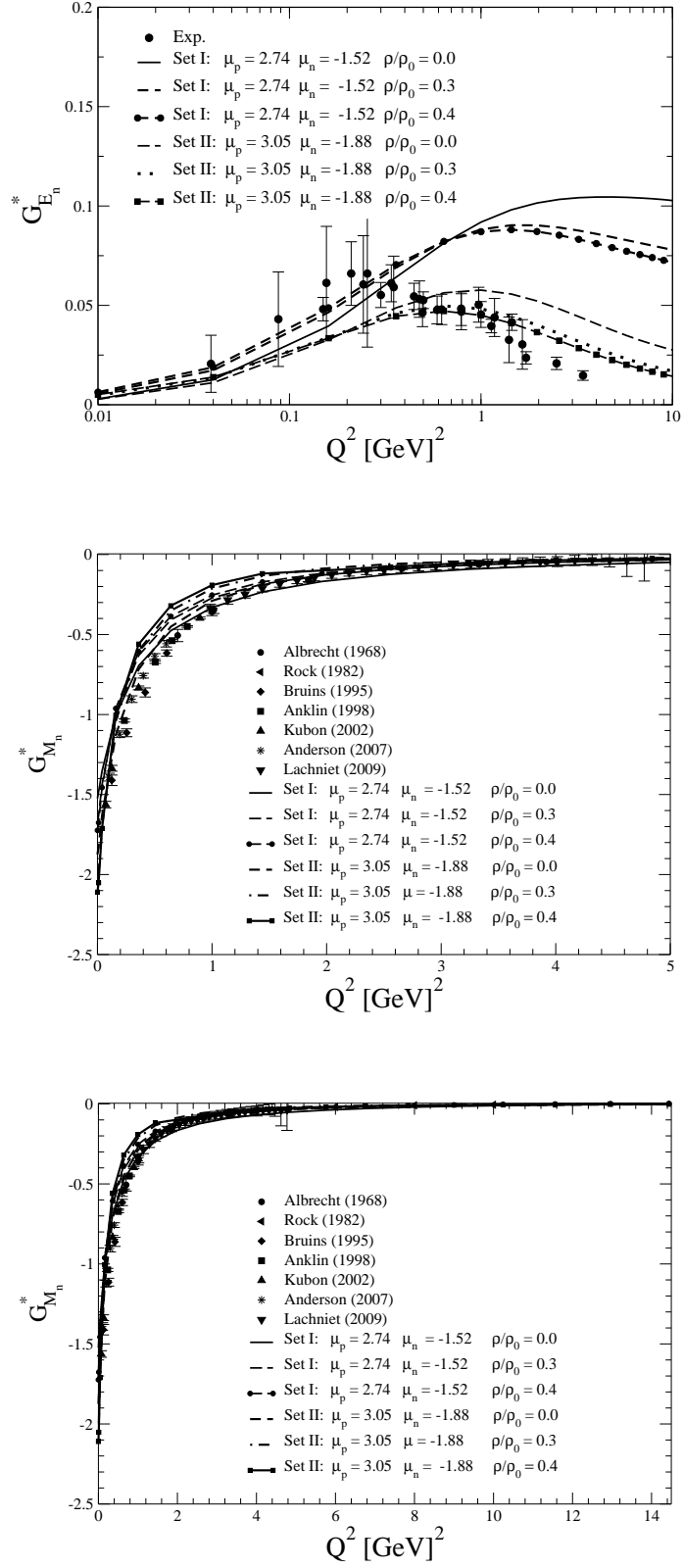

\includegraphics[scale=0.36,angle=0]{gen_mediumfig9a_new.eps}
\\
\vspace*{1.1cm}
\includegraphics[scale=0.36,angle=0]{gmnmediumfig9b_new.eps}
\\
\vspace{1.1cm}
\includegraphics[scale=0.36,angle=0]{gmnmediumfig9c_new.eps}
\caption{
Neutron electric $G^*_{En}(Q^2)$ (upper panel) and magnetic  
$G^*_{Mn}(Q^2)$ (middle and bottom panels) form factors in medium, 
obtained by the two-scale model with the two sets of parameters, 
for $\rho = 0.3\rho_0$ and $0.4\rho_0$.
Those in vacuum are shown for references.
} 
\label{gengmn_medium}
\end{figure}

This is our second main results and prediction of this article.
For each parameter set, the value of $Q^{2*}_0$ to cross zero satisfying 
$\mu^*_p G^*_{Ep}(Q^2_0)/G^*_{Mp}(Q^{2*}_0) = 0$, becomes smaller than that in vacuum.
This reflects that the in-medium falloff of $G^*_{Ep}(Q^2)$ becomes faster than 
that in vacuum as already mentioned. 
Thus, it is very interesting to pursue experiment to measure 
the proton EM form factor ratio of the bound proton, 
to check if this $Q^2$ reduction of crossing the zero 
can be observed, although such experiment would be very challenging.  
However, we would like to emphasize that this is a very interesting prediction 
of the present study. 
We believe that this is the first time prediction which is made 
with an explicit calculation.  

Next, in Fig.~\ref{gengmn_medium} we show the in-medium neutron electric  
$G^*_{En}(Q^2)$ (upper panel), and magnetic $G^*_{Mn}(Q^2)$ (lower panel) form factors.

For both parameter sates, set I and set II, 
$G^*_{En}(Q^2)$ is suppressed than that in vacuum as increasing $Q^2$, 
while very small region of $Q^2$, $G^*_{En}(Q^2)$ is enhanced than that in vacuum.

\begin{figure}[hbp]
\vspace*{6ex}
\includegraphics[scale=0.5,angle=0]{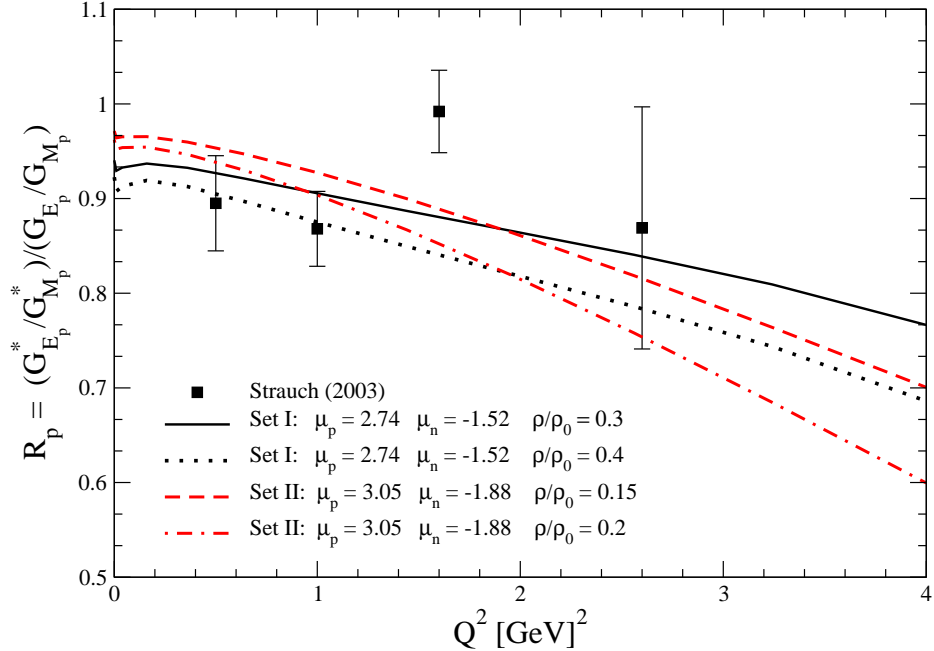}
\caption{Proton EM form factor double ratio in symmetric nuclear matter, 
$R_p \equiv [G^*_{Ep}(Q^2)/G^*_{Mp}(Q^2)]/[G_{Ep}(Q^2)/G_{Mp}(Q^2)]$, 
calculated by the two-scale model with the two parameter sets, with the set I  
for nuclear matter densities $0.30 \rho_0$ and $0.40 \rho_0$, 
and with the set II for nuclear matter densities $0.15 \rho_0$ and $0.20 \rho_0$, 
compared with the JLab data extracted for 
$[G^{^4{\rm He}}_{Ep}(Q^2)/G^{^4{\rm He}}_{Mp}(Q^2)]/[G^{^1{\rm H}}_{Ep}(Q^2)/G^{^1{\rm H}}_{Mp}(Q^2)]$.
The experimental data are taken from Ref.~\cite{Strauch1}.
} 
\label{figratio}
\end{figure}

\begin{figure}[htb]
\includegraphics[scale=0.5,angle=0]{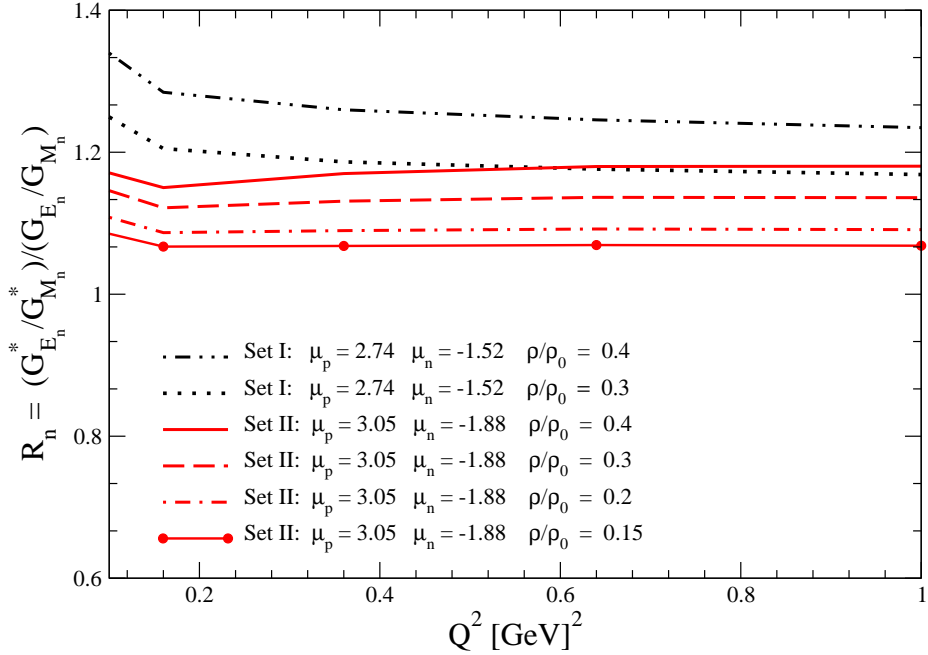}
\caption{Neutron EM form factor double ratio in symmetric nuclear matter,
$R_n \equiv [G^*_{En}(Q^2)/G^*_{Mn}(Q^2)]/[G_{En}(Q^2)/G_{Mn}(Q^2)]$, 
calculated by the two-scale model with the two parameter sets, with 
the set I and set II for nuclear matter densities $0.30 \rho_0$ and $0.40 \rho_0$, 
and with the set II for nuclear matter densities $0.15 \rho_0$ and $0.20 \rho_0$. 
The results are shown for the $Q^2$ range, $0.1 < Q^2 < 1$ GeV$^2$.  
(See also caption of Fig.~\ref{figratio}.)
} 
\label{fignratio}
\end{figure}

As for the magnetic form factor, the absolute values in medium $|G^*_{Mn}(Q^2)|$, 
becomes smaller than that in vacuum for whole region of $Q^2$ studied.
This means that the ``negative falloff'' becomes faster, or $Q^2$ dependence 
becomes more sensitive, similar trend as that found for the proton.

In Fig.~\ref{figratio},~we show our third main result of this article, 
the result of the proton form factor double ratio, 
$[G^*_{Ep}(Q^2)/G^*_{Mp}(Q^2)]/[G_{Ep}(Q^2)/G_{Mp}(Q^2)]$ in symmetric nuclear matter,   
by the two-scale model with the set I and set II, compared with the JLab data. 
The good description of the JLab data is obtained by the both parameter sets, set I and set II, 
by adjusting different nuclear matter densities.
Namely, for the set I with the nuclear matter density $0.3 \rho_0$, and for the set II 
the nuclear matter density $0.15 \rho_0$, the JLab data are well described.
Recall that the zero of $\mu_pG_{Ep}(Q^2)/G_{Mp}(Q^2)$ and the 
nucleon EM form factors in vacuum can be better described with the set II.  

We have also calculated the double ratio using the one-scale model that fits experimental 
proton magnetic moment better for the nuclear matter densities $0.3 \rho_0$ and $0.4 \rho_0$ as examples.  
But it gives a very poor description of the data, and thus we do not show the results. 
For a comparison between the one-scale and two-scale models 
made in the past, see Ref.~\cite{Araujo2006}.

Finally, similar to the proton EM form factor double ratio, 
we also show the calculated, corresponding very interesting double ratio 
for the neutron in symmetric nuclear matter $R_n$, 
where this quantity is predicted in Ref.~\cite{Cloet} to be 
enhanced in medium relative to that in vacuum for small $Q^2$ range 
($0.1 < Q^2 < 1$ GeV$^2$), while that for the proton in medium is quenched.
We show in Fig.~\ref{fignratio}
the calculated double ratios for the nuclear matter densities $\rho = 0.3 \rho_0$ 
and $0.4 \rho_0$ for the two parameter sets, 
as well as $\rho = 0.15 \rho_0$ and $0.20 \rho_0$ for the set II. 
The results show that the enhancement of the ratio in nuclear matter 
relative to that in vacuum. (Note that proton case for the densities $0.3 \rho_0$ 
and $0.4 \rho_0$ are quenched, although all the results are not shown explicitly 
except for the results shown in Fig.~\ref{figratio}.)
Thus, our results agree with the prediction made in Ref.~\cite{Cloet}.
Note that, all the double ratio $R_n$ calculated in symmetric nuclear matter 
with the densities chosen in this study, is enhanced and larger than unity 
for the $Q^2$ range $0 < Q^2 < 11$ GeV$^2$ in the present model.
The enhancement of the double ratio for neutron $R_n$ 
for $0 < Q^2 < 2$ GeV$^2$ with $0.5 \rho_0$ and $\rho_0$,   
and the quenching for the proton $R_p$ in symmetric nuclear matter, 
were also obtained in Ref.~\cite{Gilberto} by the covariant 
spectator quark model.

\section{Summary and Discussions}
\label{summary}

We have studied the nucleon electromagnetic (EM) form factors in symmetric nuclear matter  
as well as in vacuum, using a light-front motivated quark-spin 
coupling model with the one- and two-scale models of the nucleon wave functions.
The in-medium inputs for the light-flavor up and down constituent quark properties 
are obtained by the quark-meson coupling model, which has proven to be very successful 
in describing the hadron and nuclear properties in medium based on the quark degrees of freedom.

We have found that the two-scale model nucleon wave functions   
describe well the nucleon EM form factors in vacuum.
Our first prediction is that the zero of the proton EM 
form factor ratio (the zero of the proton electric form factor) in vacuum, 
to be about 15 GeV$^2$.

Based on the two-scale model with the two parameter sets which 
can respectively reproduce the proton and neutron magnetic moments reasonably well, 
we have studied the zero of the proton EM form factor 
ratio in medium. By the results, our second prediction of 
this study is that, the zero of the 
bound proton, or proton in symmetric nuclear matter, shifts to a smaller 
$Q^2$ value than that in vacuum as nuclear matter density increases.

Using the same two-scale model with the same parameter sets, 
we have calculated the proton EM form factor 
double ratio, which were extracted in JLab experiments.  
The model with the parameter set I for nuclear matter density $0.3 \rho_0$, 
and the parameter set II for nuclear matter density $0.15 \rho_0$, 
are both able to describe well the JLab data.
The results suggest that the description of the bound proton, 
or the in-medium proton EM form factor double ratio data,   
may be explained based on the internal structure 
change of the bound proton in a nuclear medium.

We have further calculated the neutron EM form factor double ratio  
in symmetric nuclear matter, corresponding to the proton case. 
Our results show that the neutron double ratio in symmetric nuclear matter 
is enhanced relative to that in vacuum, while for the proton it is quenched,  
as was theoretically predicted in Ref.~\cite{Cloet}. 
This can give an another interesting aspect to understand 
the in medium modification of the nucleon structure. 

For the future prospects, we can also study the  
nucleon axial-vector form factor in a nuclear medium with 
the same model.
Furthermore, we can extend the model to study 
the octet baryon electromagnetic and axial-vector 
form factors in vacuum, as well as those in a nuclear medium. 
\vspace{2ex}

\noindent
{\bf Acknowledgement}\\
This work was partly supported by the Funda\c c\~ao de Amparo \`a Pesquisa do Estado de
S\~ao Paulo (FAPESP), No. 2015/16295-5 (J.P.B.C.M.), and No. 2015/17234-0 (K.T.), 
and Conselho Nacional de Desenvolvimento 
Cient\'ifico e Tecnol\'ogico (CNPq), No. 401322/2014-9, No. 308025/2015-6 (J.P.B.C.M.), 
No. 400826/2014-3, and No. 308088/2015-8 (K.T.) of Brazil.

\begin{appendix}

\section{Matrix elements of the microscopic current}
\label{appendix}

The derivation of the matrix elements  of the microscopic nucleon
current operator composed by $J^+_{\beta N}$, $\beta=a,b,c,d$ of
Eq.~(\ref{mjp}) in terms of the valence nucleon wave function
follows closely Refs.~\cite{afsbw}. They are represented by the
diagrams in Fig. 1. The blobs in the figure
represent the color anti-triplet coupling of a pair of quark
fields in scalar-isocalar ($\epsilon^{lmn} \overline{\Psi}_{(l)}
i\tau _2\gamma _5\Psi_{(m)}^C$) from the effective Lagrangian of
Eq.~(\ref{lag}).

The integration over the minus-component of the momentum is   
performed to eliminate the relative light-front time in the
intermediate state propagations~\cite{sales00,tob92,Ji:1998hx}. 
This procedure allows to introduce the momentum component of the valence
light-front wave function in the computation of form factors 
(see e.g., Ref.~\cite{tob92}). 

The nucleon EM current $J^+_{N}$ derived from the
effective Lagrangian has contribution from each photo-absorption
amplitude given by the two-loop triangle diagrams of
Figs. 1a to 1d. The photon is absorbed by quark-3:
\begin{eqnarray} 
\langle s'|J^+_{aN}(q^2)|s\rangle 
&=& -m_N^2\langle N|\hat Q_q| N\rangle {\mathrm{Tr}}[i \tau_2(-i )\tau_2] \int
\frac{d^4k_1d^4k_2}{(2\pi)^8}\Lambda(k_i,p^{\prime})\Lambda(k_i,p)
\bar u(p',s')
\nonumber\\
& & \times S(k'_3)\gamma^+ S(k_3)u(p,s) {\mathrm{Tr}}\left[S(k_2)
\gamma^5 S_c(k_1)\gamma^5\right] \ ,  
\label{j+a}
\end{eqnarray}
with $\displaystyle S(p)=\frac{1}{\rlap\slash p-m+i  \epsilon} \,
,$ and $\displaystyle S_c(p)=\left[\gamma^0\gamma^2
\frac{1}{\rlap\slash p-m+i  \epsilon}\gamma^0\gamma^2\right]^T
\, $ with $T$ denoting transposition. 
The four-momentum of the virtual quark-3 after the
photo-absorption process is $k'_3=k_3+q  $. The matrix element of
the quark charge operator in isospin space is $\langle N|\hat Q_q|
N\rangle$. The function $\displaystyle \Lambda(k_i,p)$ is chosen
to introduce the momentum part of the three-quark light-front wave
function, after the integrations over $k^-$. The
contribution to the EM current represented by Fig. 
1b is given by:
\begin{eqnarray}
\langle s'|J^+_{b N}(q^2)|s\rangle 
&=& - m_N^2\langle N|\hat Q_q|
N\rangle \int \frac{d^4k_1d^4k_2}{(2\pi)^8}\Lambda(k_i,p^{\prime})
\Lambda(k_i,p) \bar u(p',s')S(k'_3)\gamma^+S(k_3)
\nonumber \\
& &\times \gamma^5 S_c(k_1)\gamma^5 S(k_2)u(p,s)
\ .  \label{j+b}
\end{eqnarray}
While the contribution to the EM current represented by
Fig. 1c is given by:
\begin{eqnarray}
\langle s'|J^+_{c N}(q^2)|s\rangle 
&=& m_N^2\langle N|\tau_2\hat
Q_q\tau_2| N\rangle \int
\frac{d^4k_1d^4k_2}{(2\pi)^8}\Lambda(k_i,p^{\prime})
\Lambda(k_i,p) \bar u(p',s')S(k_1)
\nonumber \\
& &\times \gamma^5 S_c(k_3)\gamma^+ S_c(k'_3)\gamma^5 S(k_2)u(p,s) \ .  \label{j+c}
\end{eqnarray}
Finally, the contribution to the EM current  represented by
Fig. 1d is given by:
\begin{eqnarray}
\langle s'|J^+_{d N}(q^2)|s\rangle 
&=& -m_N^2 {\mathrm{Tr}} [\hat
Q_q]\int \frac{d^4k_1d^4k_2}{(2\pi)^8}\Lambda(k_i,p^{\prime})
\Lambda(k_i,p)\bar u(p',s')S(k_2)u(p,s)
\nonumber\\
& &\times {\mathrm{Tr}}\left[ \gamma^5 S(k'_3)\gamma^+
S(k_3)\gamma^5 S_c(k_1)\right] \ .  \label{j+d}
\end{eqnarray}

The light-front coordinates are defined as $k^+=k^0+k^3\ ,
k^-=k^0-k^3 \ , \vec{k}_\perp=(k^1,k^2).$ In each term of the nucleon
current, from $J^+_{aN}$ to $J^+_{dN}$, the Cauchy integrations
over $k^-_1$ and $k^-_2$ are performed.  That means the
on-mass-shell pole of the propagators for the spectator
particles 1 and 2 of the photon absorption process are taken into
account.  In the Breit-frame with $q^+=0$, there is a maximal
suppression of the light-front Z-diagrams in $J^+$
\cite{tob92,pach99}.  Thus the components of the momentum $k^+_1$
and $k^+_2$ are bounded such that $ 0< k^+_1 < p^+$ and $0<k^+_2
<p^+-k^+_1$. The four-dimensional integrations of Eqs.~(\ref{j+a})
to (\ref{j+d}) are reduced to the three-dimensional ones on the
null-plane.

After the integrations over the light-front energies the momentum
part of the wave function is introduced into the microscopic
matrix elements of the current by the substitution
\cite{afsbw,tob92}:
\begin{eqnarray}
\frac{1}{2(2\pi)^3} \frac{\Lambda(k_i,p)}{m^2_N-M^2_0}\rightarrow
\Psi (M^2_0) \ .
\end{eqnarray}
Further, the same momentum wave function is chosen all N-q
coupling schemes for simplification. Note, that the mixed case, 
$\alpha=1/2$ in Ref.~\cite{afsbw} ($\alpha=1$ is chosen for the present 
Lagrangian density of Eq.~(\ref{lag})), 
could have different momentum dependence for
each spin coupling, however, we choose the same momentum functions
just to keep contact to the Bakamjian-Thomas (BT)~\cite{Bakamjian1953} approach.

The analytical integration of Eq.~(\ref{j+a}) of the $k^-$
components of the momenta yields:
\begin{eqnarray}
&&\langle s'|J^+_{a N}(q^2)|s\rangle = 2p^{+2}m_N^2 \langle N|\hat
Q_q|N\rangle \int \frac{d^{2} k_{1\perp} dk^{+}_1d^{2} k_{2\perp}
d k^{+}_2 }{k^+_1k^+_2k^{+\ 2}_3} \theta(p^+-k^+_1)
\theta(p^+-k^+_1-k^+_2)
\nonumber \\
&&\times {\mathrm{Tr}} \left[ (\rlap\slash k_2+m) (\rlap\slash k_1+m)\right]
\bar u(p',s')(\rlap\slash k'_3+m))\gamma^+(\rlap\slash
k_3+m)u(p,s) \Psi (M^{'2}_0) \Psi (M^2_0)
 \ ,
\label{j+alf}
\end{eqnarray}
where $k^2_1=m^2$ and $k^2_2=m^2$. The  
squared-mass of the free-three quarks is defined by:
\begin{equation}
M^2_0=p^+(\frac{k_{1\perp}^{2}+m^2}{k^+_1}+\frac{k_{2\perp}^{2}+m^2}{k^+_2}
+\frac{k_{3\perp}^{2}+m^2}{k^+_3})-{p^2_\perp} \ , \label{eqn:M0}
\end{equation}
and $M^{\prime 2}_0=M^2_0(k_3\rightarrow k'_3 \ , \vec
p_\perp\rightarrow \vec p^\prime_\perp)$.

The other terms of the nucleon current, as given by 
Eqs.~(\ref{j+b})-(\ref{j+d}) are also integrated over the $k^-$
momentum components of  particles 1 and 2 following the same steps
used to obtain Eq.~(\ref{j+alf}) from Eq.~(\ref{j+a}):
\begin{eqnarray}
&&\langle s'|J^+_{b N}(q^2)|s\rangle = p^{+2}m_N^2 \langle N|\hat Q_q|
N\rangle \int \frac{d^{2} k_{1\perp} dk^{+}_1d^{2} k_{2\perp} d
k^{+}_2 }{ k^+_1k^+_2k^{+\ 2}_3} \theta(p^+-k^+_1)
\theta(p^+-k^+_1-k^+_2) 
\nonumber \\
&&\times \bar u(p',s')(\rlap\slash k'_3+m)\gamma^+(\rlap\slash k_3+m)
 (\rlap\slash k_1+m)(\rlap\slash k_2+m)u(p,s)
\Psi (M^{'2}_0) \Psi (M^2_0) \ ,  \label{j+blf}
\end{eqnarray}
\begin{eqnarray}
&&\langle s'|J^+_{c N}(q^2)|s\rangle = p^{+2}\langle N|\tau_2\hat
Q_q \tau_2| N\rangle \int \frac{d^{2} k_{1\perp} dk^{+}_1d^{2}
k_{2\perp} d k^{+}_2 }{ k^+_1k^+_2k^{+\ 2}_3} \theta(p^+-k^+_1)
\theta(p^+-k^+_1-k^+_2) \nonumber \\
&&\times \bar u(p',s')(\rlap\slash k_1+m) (\rlap\slash k_3+m)\gamma^+
(\rlap\slash k'_3+m)(\rlap\slash k_2+m)u(p,s)
\Psi (M^{'2}_0) \Psi (M^2_0) \ ,  \label{j+clf}
\end{eqnarray}
\begin{eqnarray}
&&\langle s'|J^+_{d N}(q^2)|s\rangle = p^{+2}m_N^2{\mathrm{Tr}}[\hat
Q_q] \int \frac{d^{2} k_{1\perp} dk^{+}_1d^{2} k_{2\perp} d
k^{+}_2 }{ k^+_1k^+_2k^{+\ 2}_3} \theta(p^+-k^+_1)
\theta(p^+-k^+_1-k^+_2)
\nonumber \\
&&\times {\mathrm{Tr}}\left[ (\rlap\slash k'_3+m) \gamma^+ (\rlap\slash
k_3+m) (\rlap\slash k_1+m)\right]\bar u(p',s')(\rlap\slash k_2+m)u(p,s) \Psi (M^{'2}_0) \Psi
(M^2_0) \ .  \label{j+dlf}
\end{eqnarray}
The normalization is chosen such that the proton
charge is unity.
\end{appendix}


\end{document}